\documentclass{aa}

\bibpunct{(}{)}{;}{a}{}{,} 

\usepackage[varg]{txfonts}

\usepackage{graphicx}
\usepackage{subcaption}
\usepackage{rotating}
\usepackage[T1]{fontenc}
\usepackage{multirow}
\usepackage[]{algorithm2e}
\usepackage{ulem}

\usepackage{comment}

\begin{document}
\titlerunning{Identification of sources in gamma rays using U-shaped convolutional neural networks and a data challenge}
\title{Identification of point sources in gamma rays using U-shaped convolutional neural networks and a data challenge}

\author{Boris~Panes\inst{\ref{inst1}}\and Christopher~Eckner\inst{\ref{inst5}, \ref{inst6}}\and Luc~Hendriks\inst{\ref{inst2}}\and Sacha~Caron\inst{\ref{inst2},\ref{inst3}}\and Klaas~Dijkstra\inst{\ref{inst4}}\and Gu{\dh}laugur~J{\'o}hannesson\inst{\ref{inst7},\ref{inst8}}\and Roberto~Ruiz~de~Austri\inst{\ref{inst9}}\and Gabrijela Zaharijas\inst{\ref{inst5}} }

\institute{Pontificia Universidad Cat\'{o}lica de Chile, Avenida Vicu\~{n}a Mackenna 4860, Macul, Regi\'{o}n Metropolitana, Chile \\\email{bpanes@astro.puc.cl, bapanes@gmail.com}\label{inst1}\and Center for Astrophysics and Cosmology, University of Nova Gorica, Vipavska 13, SI-5000 Nova Gorica, Slovenia\label{inst5}\and Univ.~Grenoble Alpes, USMB, CNRS, LAPTh, F-74000 Annecy, France\label{inst6}\and High Energy Physics, Radboud University Nijmegen, Heyendaalseweg 135, 6525 AJ Nijmegen, the Netherlands\label{inst2}\and Nikhef, Science Park 105, 1098 XG Amsterdam, the Netherlands\label{inst3}\and NHL Stenden University of Applied Sciences, Professorship Computer Vision \& Data Science, Leeuwarden, the Netherlands\label{inst4}\and  Science Institute, University of Iceland, IS-107 Reykjavik, Iceland\label{inst7}\and Nordita, KTH Royal Institute of Technology and Stockholm University, Roslagstullsbacken 23, SE-106 91 Stockholm, Sweden\label{inst8}\and Instituto de F\'isica Corpuscular, IFIC-UV/CSIC, Valencia, Spain\label{inst9}}

\abstract{At GeV energies, the sky is dominated by the interstellar emission from
the Galaxy. With limited statistics and spatial resolution,  accurately separating point sources is therefore challenging.} {Here we present
the first application of deep learning based algorithms to automatically
detect and classify point sources from gamma-ray data. For concreteness we refer to this approach as AutoSourceID.} {To detect point
sources, we utilized U-shaped convolutional networks for image
segmentation and {\it k}-means for source clustering and localization. We also explored the Centroid-Net
algorithm, which is designed to find and count objects. Using two
algorithms allows for a cross check of the results, while a combination of
their results can be used to improve performance. The training data are
based on 9.5 years of exposure from The Fermi Large Area Telescope (Fermi-LAT) and we used source properties of
active galactic nuclei (AGNs) and pulsars (PSRs) from the fourth
Fermi-LAT source catalog (4FGL) in addition to several models of
background interstellar emission. The results of the localization algorithm are fed
into a classification neural network that is trained to separate the
three general source classes (AGNs, PSRs, and FAKE sources).} 
{We compared our localization algorithms qualitatively with traditional methods and
find them to have similar detection thresholds. We also demonstrate the
robustness of our source localization algorithms to modifications in the
interstellar emission models, which presents a clear advantage over
traditional methods. The classification network is able to discriminate between the three classes with typical
accuracy of $\sim$70\%, as long as balanced data sets are used in
classification training.  In this http URL\thanks{\protect\url{https://git.io/JO5FP}}, we publish our training data sets and analysis scripts and invite the community to join the data challenge aimed to improve the localization and classification of gamma-ray point sources.} {}

\maketitle

\section{Introduction}
\label{sec:introduction}

The Large Area Telescope (LAT) on board the \textit{Fermi} Gamma-Ray Space Telescope, also known as Fermi-LAT, has now been collecting photon data from energies $\lesssim 30$~MeV to more than $300$ GeV since it launched in 2008.  After more than twelve years of operation, the LAT collaboration has produced four main source catalogs, listing more than 5000 sources in the most recent one, the fourth  Fermi-LAT source catalog (4FGL), based on data collected during the first 8 years of operation~\citep{Fermi-LAT:2019yla}\footnote{An update of 4FGL, based on 10 years of operation (4FGL-DR2) is available online, at \url{https://fermi.gsfc.nasa.gov/ssc/data/access/lat/10yr_catalog/} 
}.  This large number of detected  sources, which is more than an order of magnitude improvement over pre-LAT catalogs, has made it possible to perform population studies of various  classes of gamma-ray sources and to define, for the  first time, their global properties \citep{Fermi-LAT:2PC, Fermi-LAT:4LAC}. 

The detection of (faint) point sources in the Fermi-LAT data is a challenging task, mainly due to the bright background of interstellar emission (IE), produced in the interactions of the Galactic cosmic ray population with interstellar gas and the interstellar radiation field.

Because it largely follows the structure of the gas in the Galaxy, the IE has features on all scales, the smallest of which can cause misidentification of point sources in bright regions if they are not properly modeled (see discussion on so-called {\it c} sources in \cite{Fermi-LAT:2019yla}).

For the production of the LAT catalogs, dedicated models of the IE have been developed using a maximum likelihood template fitting procedure over the whole sky \citep{Fermi-LAT:v06iem}. Point sources are then added in an iterative fashion in maximum likelihood fits of smaller regions of interest. {A point source is considered detected if the test-statistics, $TS = 2 (\log L_1 - \log L_0) > 25$, where $L_1$ is the likelihood including the point-like emission and $L_0$ the likelihood without the source.} 

Once detected, a source in the LAT catalogs is considered identified if there are clear temporal correlations with sources at other wavelengths, but associated if a statistically robust spatial association to a known gamma-ray emitter is identified in multi-wavelength catalogs. Significant differences between the angular resolution and the fields of view, as well as an uneven sky coverage of instruments covering different parts of the electromagnetic spectrum are some of the reasons that make source identification and association a challenging task. As a consequence, unidentified sources that are neither identified nor associated amount to about one third of sources in all LAT catalogs.    

While difficult, it is clear that detection and classification of faint, subthreshold sources is  critical for a range of  scientific  questions. On the one hand, numerous faint sources contain a vast amount of information on source properties and could hide yet undetected source classes. These include exotic possibilities such as dark matter subhalos as a new source population among unidentified sources (see e.g., \cite{Coronado-Blazquez:2019puc, Somalwar:2020awt} for recent progress on the topic). On the other hand, their  cumulative emission could be confused with large scale, diffuse emission and, if not accounted for, can lead to unreliable predictions of the IE and the underlying physics of the intergalactic medium, where the so-called Galactic center excess is a prime example \citep{Hooper:2010mq,Abazajian:2010zy,Hooper:2013nhl,Carlson:2014cwa,Petrovic:2014xra,Petrovic:2014uda,Mirabal:2013rba,Calore:2014xka,Bartels:2017vsx}.

The field of machine learning advanced dramatically over the past decade and {its capability for classification of unidentified point sources using the 4FGL as its input has already been demonstrated \citep{2010arXiv1007.2644M,Parkinson:2016oab,Chiaro:2016noj,Luo:2020dxa}}.  These classification works aim to guide multiwavelength searches of the unidentified sources which have proven to be a treasure trove for source finding (see e.g. \cite{2011ApJ...732...47C}), in particular for millisecond pulsars.  In this work {we explore for the first time} the potential of deep convolutional networks to both detect and classify point sources starting from the binned Fermi-LAT data of the gamma-ray sky. 

Deep learning \citep{LeCun2015_deeplearning} has demonstrated impressive results on a multitude of data sets and real-life applications like autonomous driving, voice recognition, medical applications and many more. These successes can be attributed to a couple of aspects: the scalability of these algorithms to large data sets and the development of low-cost high-performance computing hardware like Graphical Processing Units (GPUs). A particular field of deep learning revolves around Convolutional Neural Networks (CNNs), which are ideal for computer vision applications. A CNN utilizes a set of trainable convolutional layers consisting of neurons and nonlinear activation functions to learn complex patterns from data and are for example used in the case of the DeepSource algorithm \citep{Sadr:2018mud} developed for source identification in radio data, to which we return to later in the text. Also, we note that similar Deep-Learning techniques are used in \cite{Caron:2017udl} in order to analyze gamma-ray signals from the Galactic Center, which is the first time that computer vision/CNN networks are used in the context of gamma-ray analysis. The U-Net \citep{unet} architecture is a popular CNN model that uses multiscale feature extraction for learning small and large image structures.  In particular, it is powerful in addressing the task of  {\it semantic segmentation}, in which, for each pixel in the image, a class-label is predicted. The  U-Nets were originally developed for biomedical image segmentation, but have seen many other use cases, such as lunar crater detection \citep{unet_lunar_crater}, agriculture \citep{dijkstra2018centroidnet}, climatology  \citep{agrawal2019machine}, astrophysics and cosmology \citep{Caldeira:2018ojb, He_2019} or autonomous cars~\citep{Zou_2020}. 

In addition to image segmentation, deep learning has also been used for classification, as demonstrated in the seminal work \citep{NIPS2012_c399862d_deeplearning}, which sparked the immense interest in deep learning which we see today. The classification task typically has an image as input and outputs for every possible class $i$ a class probability $p_i$, where $\sum_i p_i=1$.

Motivated by  the  challenges of the traditional methods used for the gamma-ray  source detection and classification, we developed a CNN based pipeline that can be applied directly to the binned Fermi-LAT data. Using comprehensive training samples based on the source properties of the two most numerous source classes (active galactic nuclei (AGNs) and pulsars (PSRs) in 4FGL and three IE models, we trained and tested a set of U-Net based algorithms on simulated gamma-ray data to detect point sources.  Together with image segmentation networks that classify the gamma-ray images into back- and foreground on a per-pixel level, we applied the two clustering algorithms $k$-means and Centroid-Net to the segmented images to localize sources.  This allows us to compare and cross-check the algorithms, while we also explore the advantages of merging their results for improved performance.

Our classification algorithm uses as input for both training and validation the output of the source localization algorithm, {which is a small segment of the binned maps around the localized source}.  We studied the architectures and data preparation that maximize the classification accuracy.  

This work is focused on application development and uses simulated data only.  As such, we can only qualitatively compare the performance of our approach to that of the 4FGL, but can none-the-less demonstrate the potential of the method. With this document we publish 100000 gamma-ray images and additional analysis scripts, in the spirit of facilitating a community-wide data challenge, with the scope of comparing different point source finding and classification algorithms on equal grounds. 

The paper is organized as follows: In Sec. 2 we introduce the astrophysics input used for this study. In Sec. 3 we focus on point source detection and give details on training data, our algorithm and describe our performance. In Sec. 4 we focus on classification, following  the same pattern (training data, network architecture, results). In Sec. 5 we summarize and comment on future prospects.

\section{Mock gamma-ray whole sky images: Models and simulations}
\label{sec:training_data_generation}

The training data set is created by simulating the Fermi-LAT data.  It includes the two most populous source classes of AGNs and PSRs from the 4FGL (version \texttt{gll\_psc\_v20.fit} of 4FGL\footnote{\url{https://fermi.gsfc.nasa.gov/ssc/data/access/lat/8yr_catalog/}}) and three models of the IE, as detailed in the subsections below.    

\subsection{Point sources}
AGNs are a general class of bright extragalactic sources and account for 3201 identified objects in 4FGL. In this work we do not differentiate among different subclasses\footnote{\texttt{FSRQ}, \texttt{fsrq}, \texttt{BLL}, \texttt{bll}, \texttt{BCU}, \texttt{bcu}, \texttt{RDG}, \texttt{rdg}, \texttt{NLSY1}, \texttt{nlsy1}, \texttt{agn}, \texttt{sey} and \texttt{ssrq}} and include {them all} in  our data generation. PSRs are the most numerous Galactic source class, currently numbering more than 250 sources\footnote{\url{https://confluence.slac.stanford.edu/display/GLAMCOG/Public+List+of+LAT-Detected+Gamma-Ray+Pulsars}}. 

\subsubsection{AGNs}
\label{subsec:agntraining}
{\bf Spectral distribution:} We adopt a log-parabolic profile as the spectral model of AGNs (\texttt{LogParabola} in 4FGL)
\begin{equation}
\frac{\mathrm{d}F}{\mathrm{d}E} = F_{0,\mathrm{AGN}}\left(\frac{E}{E_0}\right)^{-\alpha-\beta\ln\!{\left(E/E_0\right)}}\mathrm{,}
\end{equation}
which is the most frequent spectral shape observed in {\it bright} AGNs, that is those with sufficient statistical significance to discern the true spectral shape. In order to produce synthetic AGN realizations we use the 4FGL parameters named pivot energy $E_0$ (\texttt{Pivot\_Energy}) and flux density $F_{0,\mathrm{AGN}}$ (\texttt{LP\_Flux\_Density}), together with the spectral parameters $\alpha$ (spectral slope at $E_0$, \texttt{LP\_Index}) and $\beta$ (curvature, \texttt{LP\_beta})\footnote{The label in typewriter font refers to the tag used in 4FGL.}. The energy flux $F^E_{100}$
(\texttt{Energy\_Flux100}), a quantity that we use below, is then obtained by integrating 
$EdF/dE$ for the parameters above from 0.1 to 100 GeV. We note that we also use $S_{1}$ and $S_{10}$ in the text, to mark a photon flux obtained by integrating $dF/dE$ above 1 GeV and 10 GeV, respectively. For every entry in our synthetic catalog, we draw spectral indices $\alpha$, $E_0$ and $F_{0,\mathrm{AGN}}$ from a Gaussian (log-normal for $F_{0,\mathrm{AGN}}$) whose mean and variance we determine from the respective parameters' distribution of all AGN-tagged sources in 4FGL (see Appendix~\ref{app:parameter_plots} for details). The curvature parameter $\beta$ is directly drawn from the distribution found in 4FGL without the assumption of a Gaussian-like shape. 

\noindent{\bf Spatial distribution:} To model the spatial distribution, we assume that AGNs are distributed uniformly on the sky (as expected for an extragalactic source population). Therefore, we determine the position of an individual mock object by drawing their Galactic longitude $\ell$ and $\sin$ of Galactic latitude $b$ from a uniform distribution of values between 0 and $2\pi$ and $-1$ and $1$, respectively. By drawing values of $b$ from a $\sin{b}$-distribution we guarantee a uniform distribution of sources on the surface of a sphere. 

\noindent{\bf Luminosity Function:} The number of generated sources follows the  source  luminosity function as found in the 4FGL catalog. However, since we are interested in  exploring the sensitivity  of  our algorithm to faint sources, we extend our source generation also below the 4FGL detection threshold. 

For training and verification we use a ``flat" (FL) extrapolation where we assume  that the luminosity function remains roughly constant below the Fermi-LAT detection threshold ($F^{E,TH}_{4FGL}$), extending the synthetic catalog to energy fluxes $F^{E, TH}_{1}= 3.4\times10^{-13}\;\mathrm{erg}\,\mathrm{cm}^{-2}\,\mathrm{s}^{-1}$, which is more than one order of magnitude below the 4FGL flux threshold of $F^{E, TH}_{{4FGL}}\simeq 2\times10^{-12}\;\mathrm{erg}\,\mathrm{cm}^{-2}\,\mathrm{s}^{-1}$ at high latitudes $|b|>10^{\circ}$ \citep{Fermi-LAT:2019yla}. In this case the generation of maps is rather fast, which allows us to generate  mock catalogs rather quickly. 
{For final testing of our algorithms, together with the model described above, we generate maps that adopt both a flat and a ``power-law" (PL) extrapolation,} down to a lower energy flux threshold $F^{E, TH}_{2}$ (see Fig.~\ref{fig:4FGLvsmock} left panel and a discussion on different data sets we use in Section \ref{sec:trainingdata}).

\subsubsection{Pulsars}

\noindent{\bf Spectral distribution:} Each PSR in our mock catalogs is simulated with an exponentially cut-off power law spectrum (PLEC)
\begin{equation}
\frac{\mathrm{d}F}{\mathrm{d}E} = F_{0,\mathrm{PSR}}\left(\frac{E}{E_0}\right)^{-\Gamma}\exp\!{\left[a\left(E_0^b - E^b\right)\right]}\mathrm{,}
\end{equation}
which has five free parameters: The pivot energy $E_0$ and flux density $F_{0,\mathrm{PSR}}$ as well as the spectral parameters represented by the low-energy spectral slope $\Gamma$ ($\texttt{PLEC\_Index}$), the exponential index $b$ ($\texttt{PLEC\_ Exp\_Index}$) and the exponential factor $a$ ($\texttt{PLE\_Expfactor}$) in units of $\mathrm{MeV}^{-b}$. The \textit{Fermi} Science Tools refer to this kind of spectrum as \texttt{PLSuperExpCutoff2}. Like for AGNs, the synthetic populations of PSRs is featuring randomly drawn spectral parameters according to the distribution of the respective parameters in the 4FGL. The individual distributions are again represented by a Gaussian.  

\begin{figure*}[t!]
\begin{center}
\includegraphics[width=0.48\linewidth]{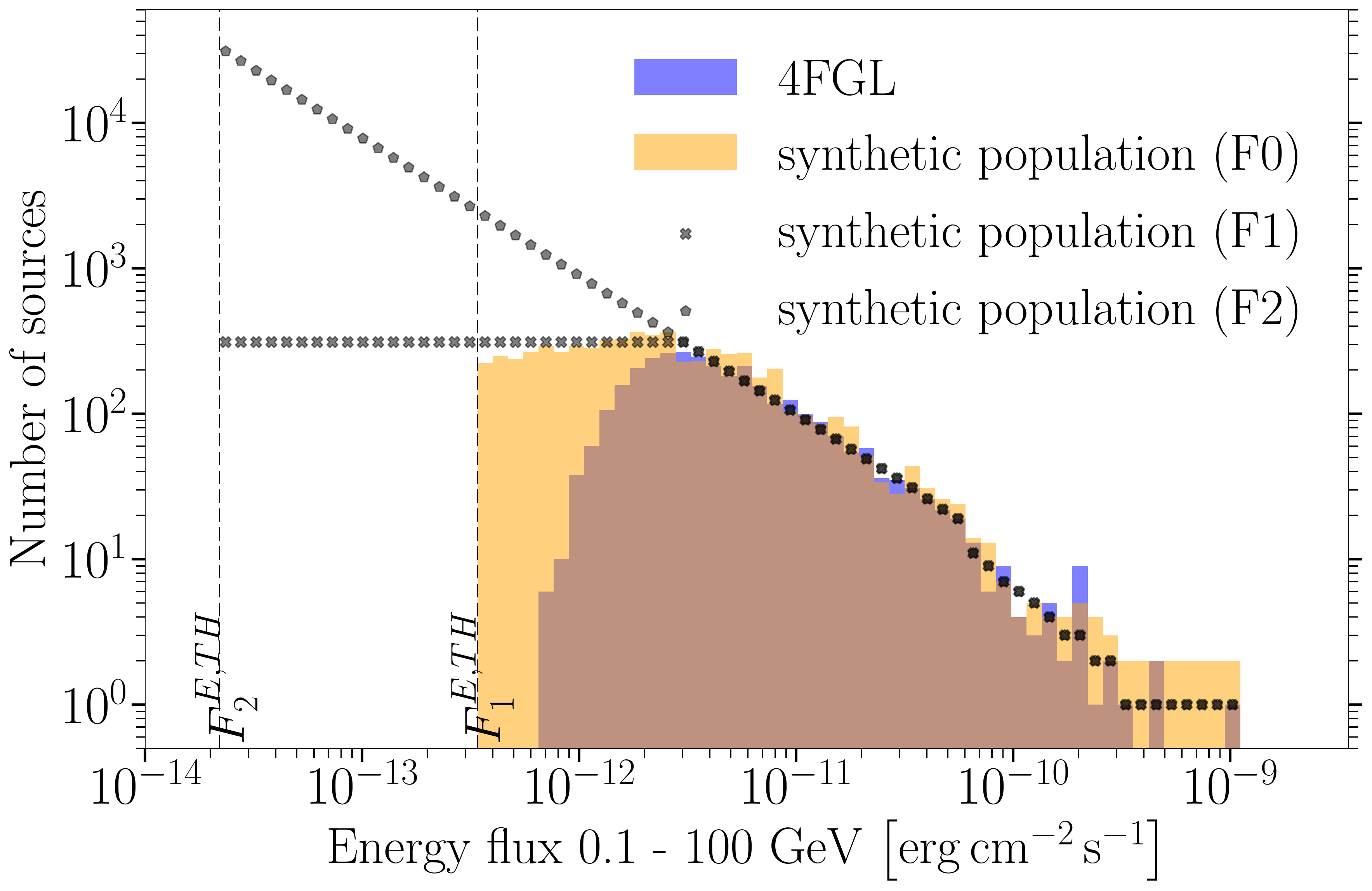}
\includegraphics[width=0.48\linewidth]{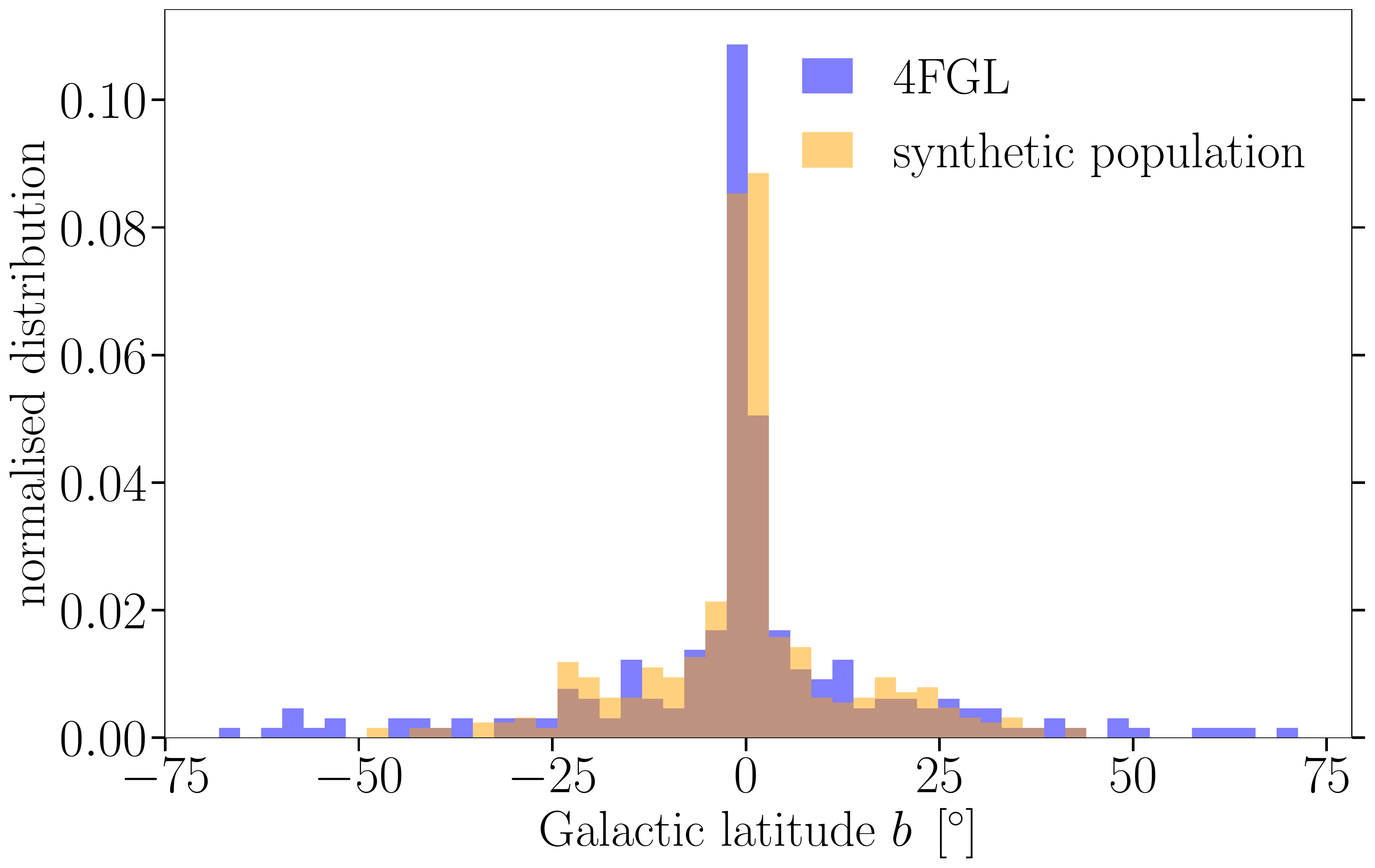}
\end{center}
\caption{\emph{Left}: Comparison of the AGN $\mathrm{d}N/\mathrm{d}F^E$-distributions described above. The blue histogram displays the actual distribution of 4FGL AGN-like sources whereas the orange histrogramme reflects the FL extrapolation to train the network, i.e.~using a the threshold $F^{E, TH}_{1}$. The distribution with gray crosses shows the FL extrapolation to the lower threshold $F^{E, TH}_{2}$, which we only use as a verification data set. The gray pentagons denote the distribution arising from the PL extrapolation. \emph{Right}: Latitude distribution of PSR sources in 4FGL (blue) and an example mock catalog of 4FGL-like PSRs (orange). We model the latitude distribution of PSRs with a double Gaussian whose parameters are given by $\left(A_1,\mu_1,\sigma_1\right) = \left(0.11, 0^{\circ}, 1.39^{\circ}\right)$ for the first Gaussian and $\left(A_2,\mu_2,\sigma_2\right) = \left(0.012, 0^{\circ}, 19.2^{\circ}\right)$ for the second Gaussian. The quantities $A_{1/2}$ are the normalization constant of the two Gaussian distributions. }
\label{fig:4FGLvsmock}
\end{figure*}

\noindent{\bf Spatial distribution:} The PSR source class contains two prominent subclasses: young PSRs which are distributed mostly in the Galactic plane and millisecond PSRs (MSPs) which migrate further from the plane, resulting in a  broader distribution. We  therefore model the latitude distribution of PSRs with a double Gaussian whose parameters are obtained by a least-squares fit, matching well the distribution  of detected sources (see Fig. \ref{fig:4FGLvsmock})\footnote{We use the {\it detected} gamma-ray PSR distribution as the true (input) one, which  is not consistent with our approach for AGNs. Since the true PSR distribution is poorly constrained we none-the-less adopt this approach which is subject to observational biases.}. The subclass residing well within the Galactic disk is modeled by a Gaussian with mean $\mu_1 = 0^{\circ}$ and a standard deviation of $\sigma_1 = 1.39^{\circ}$, while the broader distribution of migrated PSRs is given by $\mu_2 = 0^{\circ}$ and $\sigma_2 = 19.2^{\circ}$. To ensure a normalization to unity the sum of both Gaussian modes requires the normalization constants $A_1 = 0.11$ and $A_2 = 0.012$. We assume the longitude distribution of PSRs is uniform.  

\noindent{\bf Luminosity Function:} To derive a viable synthetic PSR catalog that resembles the properties of the already detected 4FGL objects, we follow the exact same general  algorithm as in the case of AGNs, using the flat extrapolation for training and both the {\it flat} and {\it power  law} extrapolation to an even lower threshold for verification (see more in Sec. \ref{sec:trainingdata}). 

\subsection{Interstellar emission models}
\label{sec:de}
As mentioned in the introduction, there is a significant level of degeneracy between the IEM and the point source detection, since for example, imperfect IEM can result in spurious source detection, or faint sources could be obscured by an overly bright IEM. The biggest uncertainty in the construction of IEMs is in the morphology of the gas emission, as the emission that correlates with gas inherits small scale features (from the varying gas density).   

As the benchmark IEM in this work, we use the latest Fermi-LAT background model, $\rm{gll\_iem\_v07.fits}$\footnote{\url{https://fermi.gsfc.nasa.gov/ssc/data/access/lat/BackgroundModels.html}}. To account for emission from unresolved extra galactic
emission and irreducible CR background in the data, we use the official isotropic template\footnote{\texttt{iso\_P8R2\_ULTRACLEANVETO\_V6\_v06.txt}}. This combination was used to derive the
4FGL catalog and we refer to it as {\bf B1}. 

To check the robustness of our method to the uncertainties in the IEM, we test our algorithms with two additional models, {\bf B2} and {\bf B3}, which were built using different methods and employ different assumptions on the galactic gas distribution.  B2 is  chosen among the models used to estimate the systematical uncertainty in the first Fermi-LAT supernova remnant (SNR) catalog, based on the SNR CR source distribution, a halo height of 4~kpc and spin temperature of 150~K \citep{Acero:2015prw}.  For B3 we chose $\rm{gll\_iem\_v06.fits}$, the model  that was built for the generation of the 3FGL catalog, but scaled for the P8R2 data \citep{Fermi-LAT:v06iem}. 

For details on the assumptions and methods used for building these three IEMs, we refer to the relevant publications and documentation.  Here we only summarize their difference.  Model B1 is the most recent of the models and uses different surveys for the distribution of gas in the Galaxy compared to models B2 and B3 that use the same surveys, affecting the fine scale structure of the models.  Models B1 and B3 were both designed for the generation of Fermi-LAT catalogs and contain a component built using filtered data-residuals.  Model B2 does not contain such a component.  Finally, while the models have all been tuned to match the Fermi-LAT data, their tuning methods are all different.  So even if models B2 and B3 use a linear combination of similar components, the final results are quite different. We complement B2 and B3 with the same IGRB model, as above.  

\subsection{Mock whole-sky data generation}
\label{sec:data}

The sky maps of our mock catalogs have been obtained via the Fermi-LAT Science Tools package version 11-07-00 which is available from the \textit{Fermi} Science Support Center\footnote{\url{https://fermi.gsfc.nasa.gov/ssc/data/analysis/}}. We used the \texttt{P8R2\_ULTRACLEANVETO\_V6}  instrument response function and applied further selection quality cuts \texttt{(DATA\_QUAL == 1) \&\& (LAT\_CONFIG == 1) \&\& (IN\_SAA != T)} to work with \texttt{FRONT+BACK} type events.  
To prepare the required Fermi-LAT exposure files, we have used 495 weeks of Fermi-LAT all-sky data taken during the time period from the 4th of August 2008 to the 1st of February 2018, corresponding to about 9.5 years of data. 

Using the corresponding spacecraft files, we derived a binned exposure map in the HEALPix format \citep{2005ApJ...622..759G} to generate all-sky maps of each AGN or PSR mock catalog and IEM models with the Science Tools routine \texttt{gtmodel}. The data were binned in five energy bins ($0.5-1$, $1-2$, $2-7$, $7-20$, $20-200$ GeV) and has a spatial resolution of $\sim0.23^{\circ}$ (RING ordering scheme with $N_{\mathrm{side}} = 256$ HEALPix maps). 

We note that this resolution is worse than the LAT angular resolution at higher  energy bins,  and therefore our results do not  exploit the full potential of the data in that respect. 

The final sky maps  are  obtained using the \texttt{gtmodel} routine and  contain photon counts that represent an infinite-statistics scenario (i.e. the expected counts if one would average over infinitely many Poisson realizations). This dataset is sometimes referred to as Asimov dataset \citep{2011EPJC...71.1554C} and we use it because it is much faster than simulating the real photon data using \texttt{gtobssim}. To each map we then add Poisson noise {\it a posteriori}, in order to account for the Poisson statistics of the real data. 

\begin{figure*}
\begin{center}
\includegraphics[width=0.98\linewidth]{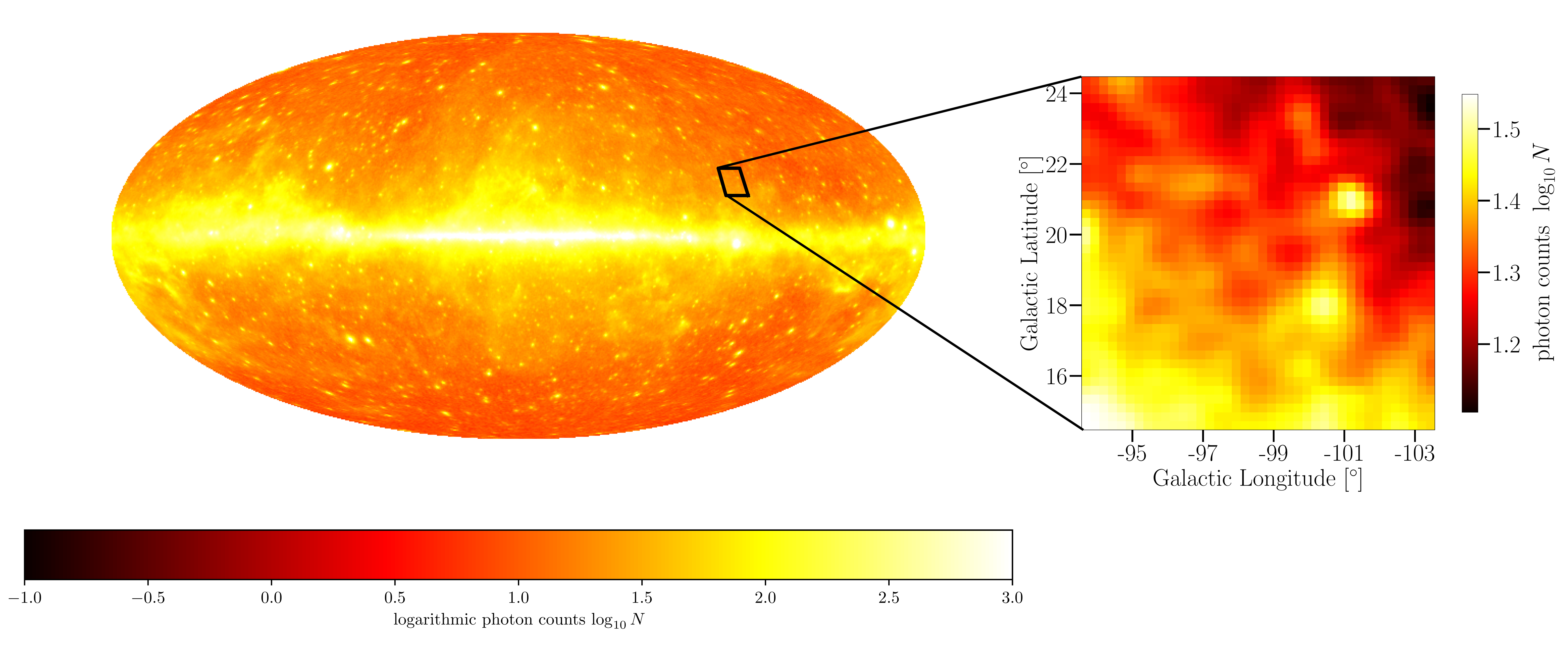}
\end{center}
\caption{Fermi-LAT gamma-ray map using data with $E>1\,$GeV from the time period and selection criteria stated in the text.  The figure uses Galactic coordinates and Mollweide projection with the Galactic center in the middle. The rectangle and zoom displays an example $10^\circ \times 10^\circ$ patch that we use as input to the detection algorithm.  The sky is evenly distributed into these patches based on the HEALPix pixelization scheme.  The all-sky and zoom-in image have been smoothed with a $1^{\circ}$ Gaussian kernel. \label{fig:LAT_sky_with_patches}}
\end{figure*}

\section{Point source finding using deep convolutional networks}
\label{mlmethods}

\subsection{Training and test data for localization algorithms}
\label{sec:trainingdata}

Each whole sky map in our training data set is a sum of the AGN,  PSR, IEM, and IGRB maps. In particular, the AGN and PSR maps are  prepared assuming the flat luminosity extrapolation to the $F^{E, TH}_{1}$  flux threshold and the IEM model used in training is B1 (see Section \ref{sec:training_data_generation}). We followed the luminosity function of identified AGNs and PSRs and therefore underpredict the number of sources on the sky (i.e. we do not simulate sources classified as unassociated in  the  catalog). In addition, we do not make an effort to balance the data set, in terms of the relative numbers of PSRs and AGNs. We return to this issue when discussing the training set for classification, in Section \ref{sec:classification}.  

To reduce the dimensions of the input data and simplify our  training procedure, we split the whole sky maps into $10^\circ \times 10^\circ$ ROI ('patches') corresponding to $64\times64\times5$ pixels per ROI. For training, {each sky map instance contributes with 400 patches}, with positions chosen from a flat distribution in $l$ and $\sin b$, where $l$ and $b$ are Galactic longitude and latitude. For testing, we instead uniformly distributed the patches to cover the whole sky, using 768 patches based on a HEALPix pixelization with $N_{\mathrm{side}}=8$. In both cases, patches are then projected into Cartesian coordinates (see Fig.~\ref{fig:LAT_sky_with_patches})\footnote{Obtained using the healpy projection functionality.}.   

Since we use simulated data for the training of the algorithm, in principle we are able to generate training data samples that contain as many patches as we want, as long as we consider independent sky map instances. Thus, in order to determine and fix the training and validation sample size which is used to report the results shown in this work, we considered the evolution of the performance of our algorithm as a function of the sample size. In practice, we produced five samples of mock data, containing 1000, 5000, 10000, 50000, and 100000 patches respectively. Later, we computed the performance of the localization algorithm in terms of {\it Purity} (see Eq.~\ref{eq:purity}) and {\it Completeness} (see Eq.~\ref{eq:completness}) metrics considering every network trained with these samples. 

The results demonstrates that {\it Completeness} has converged to a stable value. While {\it Purity} is still increasing, we decided that 100000 patches is a good compromise between computational time and {\it Purity}. We also note that we have a classification pipeline applied after the localization step which is capable of increasing the {\it Purity} of the sample. And for the training of the classification step, it is actually beneficial to have some examples of false positive (fake) sources.


During the training of the detection network, we divided each of these data samples into independent subsets containing $\sim 80\%$ and $\sim 20\%$, respectively, for training and validation, reserving one map  for testing. Besides, we generate plain text files (CSV format) with the list of patches and source positions. These files are useful to efficiently record the true information of simulated patches. Also, the output of localization and classification algorithms follow a similar format in order to simplify the process of model evaluation. 
 
We test the robustness of our algorithm using three independent point source maps. The first one is defined as {\bf F0}, which is a single point source whole sky map, based on the flat extrapolation down to the low flux thresholds of $F^{E, TH}_{1}$ (i.e. the same point-source generation as in the training set). The second one is given by {\bf F1}, which is a single point source set that is obtained by combining flat luminosity function extrapolation for AGNs and PSRs, down to the lower flux threshold $F^{E, TH}_{2}$. Finally, the third map is defined as {\bf F2}, which is a single point source set that is obtained by  combining the {\it power-law} luminosity function extrapolation for AGNs and PSRs down to $F^{E, TH}_{2}$ (see Fig.~\ref{fig:4FGLvsmock}). Each of these three point  source sets is then combined with all three IEMs, namely {\bf B1}, {\bf B2} and {\bf B3} for the total of nine independent sky maps, which we use to test the network performance.    


\subsection{Network structure}

In order to model a predictor of point source positions, which is able to work directly from patch images of the sky, we considered two machine learning algorithms which are based on the U-Net architecture~\citep{unet}. This algorithm is able to produce relevant information about the per-pixel classification of background (diffuse gamma-ray sky) and foreground (point) sources. However, a per-pixel classification of point sources is not yet a point source detection and one needs to cluster the pixels together. Thus, in the second step of our analysis we used two different approaches for automatic clustering and prediction of point source positions (see Fig.~\ref{fig:temp-pipeline}). The first one is based on the traditional $k$-means clustering algorithm. In the second approach we used Centroid-Net, which is an algorithm that integrates the U-Net (classification and regression modes) with a Voting Algorithm in order to produce robust predictions. Below, we discuss the U-Net basic architecture and both algorithms for clustering.

It is worth stressing that, while the input data (left-most image in Fig.~\ref{fig:temp-pipeline}) contains the information on the Fermi-LAT response functions, in particular its Point Spread Function  (PSF) in five energy bins, the second, clustering input data (central column images in Fig.~\ref{fig:temp-pipeline}) is defined on a region with a fixed single dimension (radius or side), which is a hyper-parameter chosen in order to optimize the clustering algorithm. 
 The parameter chosen here is a compromise between the sizes of all energy bins and in combination with the fixed spatial binning limits the location accuracy.  A more optimized approach where the size of the source follows that of the PSF will be considered in future work.

\subsubsection{U-Net plus k-means (UNEK)}

\begin{figure*}[ht]
\begin{center}
\includegraphics[scale=0.35]{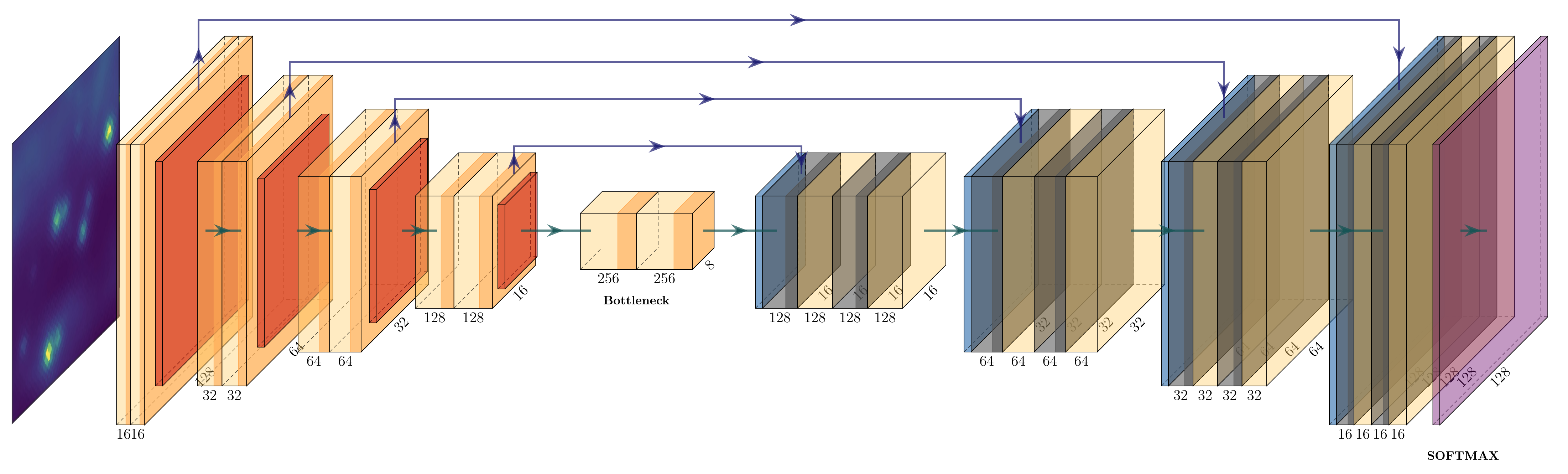}
\end{center}
\caption{Visualization of the U-Net architecture as implemented in this work (see  text for details).}
\label{fig:unet}
\end{figure*}

The U-Net~\citep{unet} is a specific FCNN\footnote{A CNN without dense layers} architecture, where an $(w,h,c)$-shaped input image is transformed in a $(w',h',c')$-shaped output image. Here, $w$ and $h$ are the input image width and height and $c$ is the number of channels (in a color image the number of channels would be 3: one for green, red and blue). The output of the U-Net is also an image, but the dimensions and number of channels can change (see Fig.~\ref{fig:unet}). 

The U-Net architecture works as follows. After every layer of the initial part, the image dimension is halved while the number of channels is increased. In the middle layer (the bottleneck), the image is very small and has a very high number of channels. Next, the image dimension is increased again until it has the target image dimensions. The U-Net needs to compress as much information as possible in this tiny image, leading to high information density on a small scale. Meanwhile, every convolutional layer is also connected to its equally-sized counterpart after the bottleneck. This means the small scale structure and the large scale structure are combined to make accurate small scale predictions on the large scale image map. 

For the particular problem of point source localization, the U-Net can be used to produce segmented regions around point source centers or vector positions that point to the centers of each source. It  is necessary then to apply a clustering algorithm to synthesize this information in terms of point source positions.

In our first algorithm, called UNEK, we trained the U-Net in order to produce segmented regions around point sources and afterwards we applied a $k$-means based algorithm for clustering and point source position identification. We also adopted an independent approach called CNET that  we  describe  in the next section. 

\begin{figure*}[ht]
\begin{center}
\includegraphics[scale=0.4]{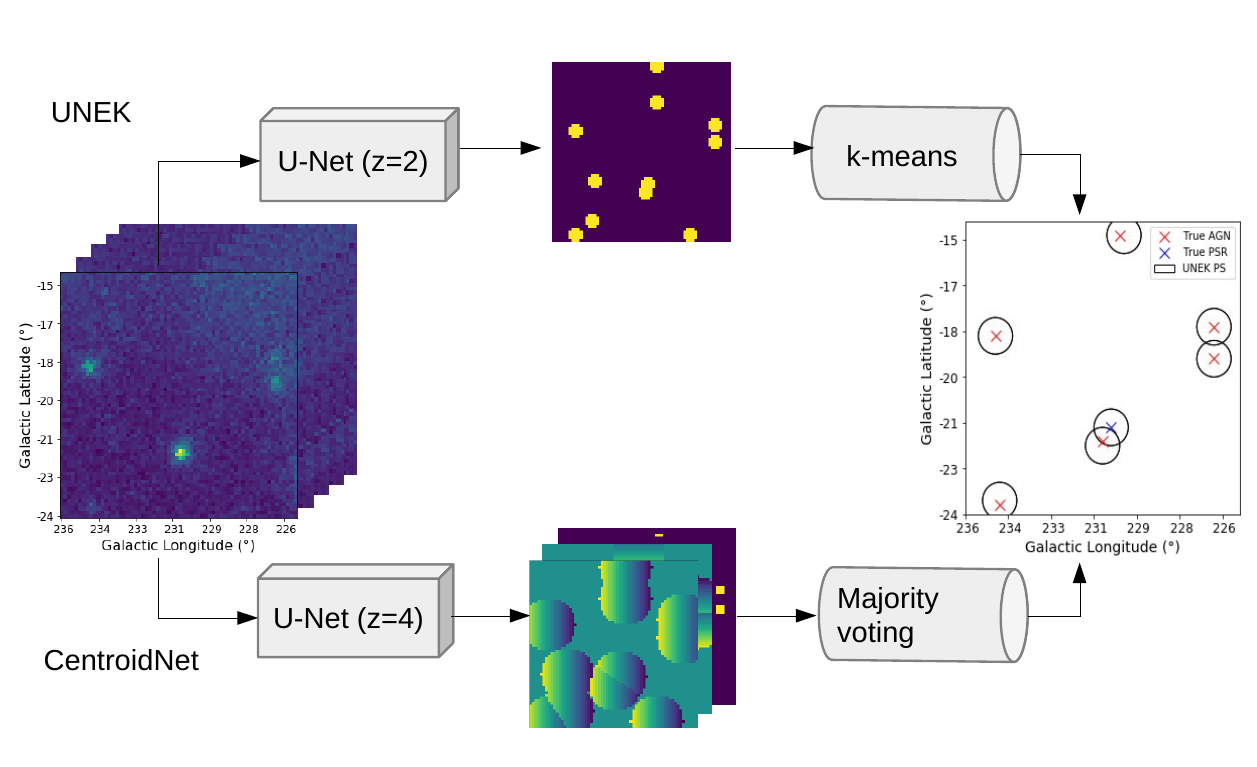}
\end{center}
\caption{The panel depicts the flux diagram for both localization algorithms discussed in this paper, UNEK and Centroid-Net. Both algorithms receive as input gamma-ray sky patches with resolution (64,64,5), then the U-Net is applied in order to obtain semantic information about the localization of sources. Finally, each algorithm uses a particular clustering algorithm to recover the position of sources. Note that the iterative k-means algorithm uses  only one channel as an input, which majority voting algorithm uses three layers (see text for details).}
\label{fig:temp-pipeline}
\end{figure*}

In the case of $k$-means, the U-Net target contains two maps: a map that is 0 on all background and 1 on all foreground (the point sources), and the inverse of this. In practice, we assumed that point source segmented regions are given by solid disks around each point source position, with a fixed and constant radius $R$. The radius of these regions must be fixed in advance, and we found that $R=2.5$ pixels gives good results. The U-Net flux diagram is shown in the top panel of Fig.~\ref{fig:temp-pipeline}, together with the image representation of input and output data.

The network is trained using 50 epochs with a batch size of 128 patches. The input images had a dimensionality of $64\times64$ and 5 channels. We used reduce learning rate on plateau, with patience of 5 epochs, reduce factor of 0.1 and minimum learning rate of $10^{-5}$. We used checkpoints to save only the best model and we considered a training setup that includes validation data loss evaluation at each epoch in order to check/avoid overfitting. 

The expected output of the U-Net (first layer) for each input image patch is given by a set of label scores (softmax outputs) that are close to 1 for every pixel that belongs to the circular regions around the center of each point source position and close to 0 otherwise. In the cases used for the performance study of UNEK, we obtain a U-Net binary accuracy (comparison between target and predicted layers using a threshold of 0.5 in order to assign 1 or 0 to the U-Net output) around 0.96, which indicates that the U-Net predictions indeed follow these expectations.

At this point we can apply $k$-means in order to cluster the well defined circular regions predicted by the U-Net and compute the position of the most stable centroids (point source positions). In $k$-means, the positions of the clusters are discovered by a process that aim to minimize the sum of the square euclidean distance between each vector to the center of the nearest cluster. By definition, the $k$-means algorithm starts with the random generation of $k$ centroids (cluster positions) and during the optimization process the number of clusters is fixed. Thus, in order to run $k$-means we need to know the number of clusters to look for in advance. If for example there are five point sources on a particular image patch, one could do a $k$-means clustering method with $k=5$ to determine the point source locations.  The $k$-means optimization would return the centers of the five point sources. However, \textit{a priori} we do not know the number of point sources. 

Therefore, the target of the following $k$-means based clustering algorithm is to automatize the process of point source localization considering as input data only the semantic information contained in the U-Net output. In particular, the algorithm must predict the number of sources as part of the process. The expected flux diagram for this algorithm is shown in Fig.~\ref{fig:temp-pipeline}. Below, we describe one functional prototype that executes this action (see  Appendix \ref{app:UNEKalgorithm} for more details). 

As an input to  $k$-means, we only considered one single layer of the U-Net output, which contains the segmented fixed radius disks at the positions of point sources. This single layer is the result obtained from the processing of the full input image (see upper top panel of the central column of Fig.~\ref{fig:temp-pipeline}). From the whole set of pixels, we focused only on those with a label score that overpass a given threshold. In practice we have set this value to 0.2 which proved to give reliable results, though a systematic study  might be merited. Thus, if the U-Net output is accurate, we end  up with a set of pixels which are naturally grouped around the position of sources. Considering these pixel positions we can generate the (x,y) vectors that we use in order to feed $k$-means.

To continue, we note that, for a fixed value of $k$, the cluster centers predicted by $k$-means can be used to generate a 2D image that contain solid fixed radius circular regions around the centers of these clusters, just like the U-Net expected output. Let us define this image as $I_k$. Thus, if we chose the radius of $I_k$ disks as $R$, we can directly compare $I_k$ to the expected output of the U-Net, pixel by pixel. At least intuitively speaking, the overlap between these images should be maximized by a value of $k$ close to the true number of sources. 

In practice, we can indirectly quantify the overlap between $I_k$ and the U-Net output by considering the U-Net predictions. In order to compute this effective overlap, defined as $S_k$, we considered the sequential visit to the predicted circular regions of $I_k$. At each step, we updated the value of $S_k$ considering the sum of the scores inside each region. The value of $S_k$ should increase with the value of $k$, for $k$ lower that the number of true sources $N_s$, since the covered area is increasing and there are available high score pixels. On the other hand, when $k>N_s$ the circular regions in $I_k$ would start to share some pixels, since the new $k$-means clusters only can arise as divisions of previous ones. Thus, each time that a pixel is included in more than one region, we penalized the value of $S_k$ by a negative amount. Therefore, in optimal conditions the value of $S_k$ would be maximized when $k$ approaches to $N_s$. The pseudo code of this algorithm is given in the Alg.~\ref{alg:kmeans-based} in the Appendix.

\subsubsection{Centroid-Net (CNET)}

In Centroid-Net the output of the U-Net model (which is an integral part of the Centroid-Net package\footnote{https://github.com/kdijkstra13/OpenCentroidNet}) is redefined to produce robust centroids of objects. In addition to the class-probability outputs of U-Net, two output layers are added to the network, which are then trained to predict a per-pixel voting vector that point to the nearest centroid of an object, see Fig.~\ref{fig:temp-pipeline}. Instead of grouping pixels to produce meaningful objects like in the UNEK approach, the voting vector field of Centroid-Net is decoded to directly produce centroids of objects. The amount of votes that is received for each pixel is counted and locations with a high number of votes are regarded as object centroids. This approach has several advantages for detecting point sources: firstly, each image location contributes to the localization of point sources, not only the foreground pixels as with the UNEK approach, secondly, a more robust detection could be achieved because of the majority-voting scheme and thirdly, the voting-output space can be visualized and post processed to further analyze the performance of the algorithm and mitigate the black box nature of neural networks.

We found that the optimal results are obtained using 300 epochs for training with a batch size of 100 using the Adam optimizer with a learning rate of $10^{-3}$. {For the hyper-parameters of the voting algorithm (see \cite{dijkstra2018centroidnet} for details) we have set the amount of spatial binning to 1, the distance between two centroids should be smaller than 1, and that 10 votes constitute a centroid.}

\subsection{Performance parameters}

To quantify the performance of our localization algorithms, we followed the example of the DeepSource project (a deep learning package that uses convolutional neural networks for radio source  detection \citep{Sadr:2018mud}). Performance there is quantified via {\it Purity} (known also as a {Precision}  in  the ML community) and {\it Completeness} (i.e. Recall, or Efficiency~\citep{DiMauro:2017ing}).

Purity is defined as the fraction of the total number of detected sources that are real:
\begin{equation} \label{eq:purity}
Purity = \frac{\textrm{TP}}{\textrm{TP} + \textrm{FP}}
\end{equation}

\noindent where {\bf true positives (TP)}, is the number of matched pairs and {\bf false positives (FP)} is the number of sources  detected by the algorithm that do not possess a counterpart in the true, synthetic point source catalog. Both of these quantities are functions of flux or, alternatively, a signal-to-noise ratio (SNR$_c$), where SNR$_c$ is computed as the ratio between the amount of photons from the point source (signal) over the square root of the background plus signal photons (total counts), both evaluated at the location of the point source (central pixel). 

{Completeness} is defined as the fraction of true  sources  that  are  successfully  recovered  by  the  algorithm, that is 
  
\begin{equation} \label{eq:completness}
Completeness = \frac{\textrm{TP}}{\textrm{TP} + \textrm{FN}}
\end{equation}

\noindent where {\bf false negatives (FN)} is the number of true point sources not found by the algorithm. 

 We calculated the uncertainty on {Purity and Completeness parameters using the standard expressions for uncertainty propagation, for which we assume that the number of TP, FP and FN distribute as Poisson variables. For instance, if we use $p$=Purity and $N$=TP+FP, the respective uncertainty of $p$ would be given as $\sqrt{p(1-p)/N}$.}

Another metric, that is used to gauge the ability  of an algorithm to detect faint sources is {\it S90}, defined as the lowest SNR$_c$ (or flux) above which an algorithm produces both Purity and Completeness values of at least 0.9.

We defined that the source found  by the algorithm is {\it matched} with a true source if the distance between them is smaller than 0.3 degrees\footnote{In comparison, almost all sources in the 4FGL catalog are localized within 10 arcmin with most having accuracy of around 3 to 5 arcmin (see Fig.~18 in \cite{Fermi-LAT:2019yla}). We note that the \textit{Fermi}-LAT $68\%$ photon confinement region (angular resolution) for FRONT events of the \textit{Fermi} LAT ranges between 0.4$^{\circ}$ at 1 GeV to 0.1$^{\circ}$, for $>10$ GeV, \url{https://www.slac.stanford.edu/exp/glast/groups/canda/lat_Performance.htm}}. In case there is more than one true source found within 0.3 deg, we associated the detected source with the true source that has the highest photon flux value, as the most likely counterpart. {For TP sources we used the photon flux of the associated source in the performance plots below. For FP sources -- since there is no true source associated to the predicted one -- we used the flux of the closest true source.}

\subsection{Results of PS finding algorithm}

\begin{figure*}[h]
\begin{center}
\includegraphics[scale=0.5]{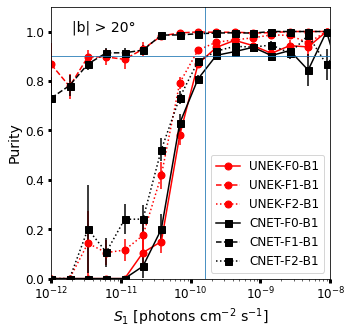}
\includegraphics[scale=0.5]{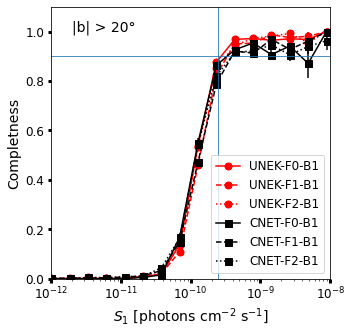}
\includegraphics[scale=0.5]{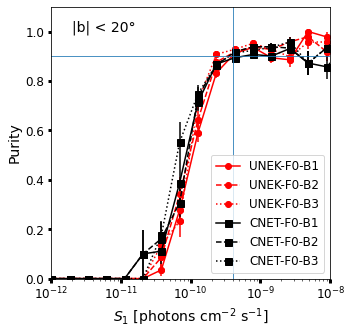}
\includegraphics[scale=0.5]{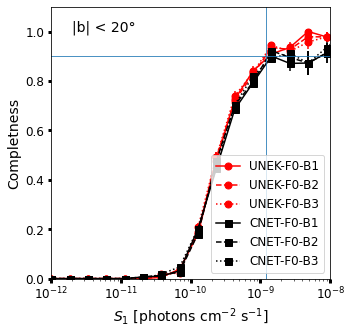}
\end{center}
\caption{Purity (left) and Completness (right) as a function of photon flux  above 1 GeV, $S_1$. The top panel focuses on for high latitudes and shows {results for the three different luminosity functions (F0-F2 test data)}, while the bottom pannel focuses on low latitudes and shows the {results for the three different IEM  (B1-B3  models)}.  Results of  the UNEK algorithm are shown in red, and CNET in black.}
\label{fig:temp}
\end{figure*}

\begin{figure*}[h]
\begin{center}
\includegraphics[scale=0.5]{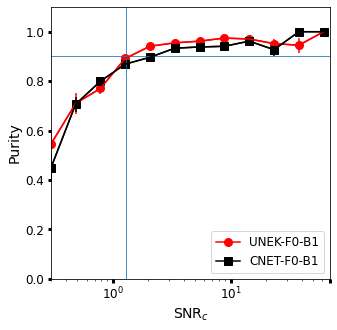}
\includegraphics[scale=0.5]{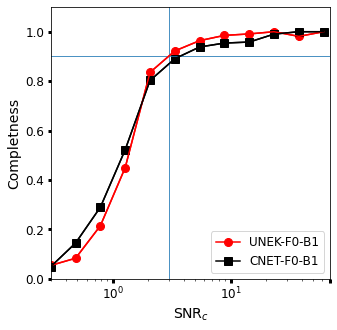}\\
\includegraphics[scale=0.5]{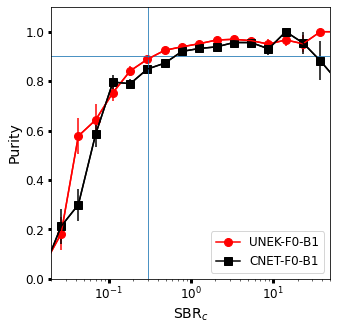}
\includegraphics[scale=0.5]{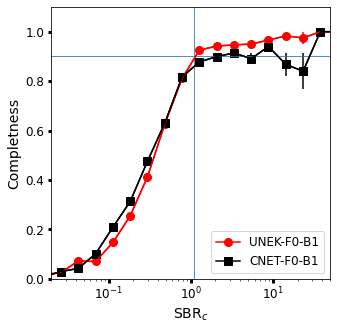}
\end{center}
\caption{Purity (left) and completness (right) as a function of Signal-to-Noise ratio (top) and Signal-to-Background ratio  (bottom).}
\label{fig:tempb}
\end{figure*}

Our results, shown in Figs. \ref{fig:temp} and \ref{fig:tempb}, can  be summarized as follows: 

\begin{itemize}
    \item {\bf Top} panel of Fig.~\ref{fig:temp} focuses on high latitudes $b>20^\circ$, and shows Purity and Completeness as  a function of the source photon flux $S_1$ (see Sec.~\ref{subsec:agntraining}). We show the performance of both of our algorithms, UNEK (red) and CNET (black) for the F0 test catalog (solid), F1 (dashed) and F2 (dotted) catalogs. We observe that results  are quite independent of the catalog choice (i.e.~are robust regarding the number of sub-threshold sources in  the mock data). In terms of the network architecture, CNET and UNEK  seem to perform comparably well. 
    
    \item {\bf Bottom} panel of Fig.~\ref{fig:temp} focuses on low latitudes ($b<20^\circ$) and shows Purity and Completeness as a function of the source photon flux $S_1$, for the F0 catalog,  obtained by the UNEK (red) and CNET (black) algorithms. In addition to the baseline  IEM, B1 (solid), we also show the results on the test data that use alternative background models (B2-dashed and B3-dotted). We note that even though the network was trained on B1, its  performance {(in terms of source numbers)} remains very similar, also when other models are used. 
   
   \item In Fig.~\ref{fig:tempb}, {\bf Top}, we show Purity and Completeness for our benchmark case  (UNEK, F0, B1) as a function of SNR$_c$ for all latitudes. {The behavior of Purity with the SNR$_c$ is as expected, but the Completeness shows we are background dominated with many high SNR$_c$ sources going undetected.}
   
   \item {To further demonstrate the background dominance}, in Fig.~\ref{fig:tempb}, {\bf Bottom}, we also show our performance as a function of signal-to-background ratio (SBR$_c$). SBR$_c$ is defined as the ratio between the amount of photons from the point source (signal) over the background photons (that include diffuse emission and possible overlapping sources). {As expected, the Completeness drops quickly when the SBR goes below 1, however Purity remains above $90\%$ down to SBR=0.3.}

\end{itemize}

{While the global properties of the results from UNEK and CNET shown in Figs.~\ref{fig:temp} and~\ref{fig:tempb} are nearly identical, the results differ in detail and there is not a perfect overlap between the TP and FP sources in the results.  A comparison between the results of the UNEK and CNET algorithm for the F0-B1, F0-B2, and F0-B3 simulations show that around 90\% of the TP sources in the results are identical between the two algorithms, while only 10\% of the FP sources are identical.  This can be used to our advantage to increase either the Purity or Completeness of the signal.  By taking the union of the two results, the number of TP sources are increased by roughly 10\%, at the expense of nearly doubling the number of FP sources.  This can push the Completeness threshold down a bit, at the expense of worse Purity.  Using instead the intersection of the two results, the number of FP sources can be reduced by 90\% with only a 10\% decrease in the TP sources, resulting in a much higher Purity at the expense of Completeness.
}

\subsection{Comparison with the traditional LAT catalogs} 

\begin{figure*}[h]
\begin{center}
\includegraphics[scale=0.5]{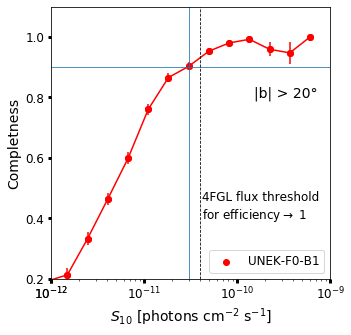}
\includegraphics[scale=0.5]{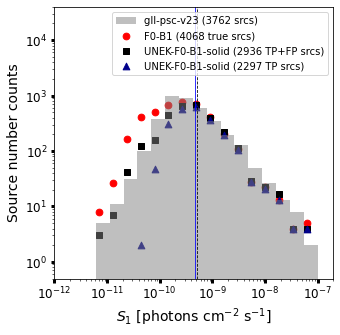}
\end{center}
\caption{{{\it Left:} Completness of our algorithm at high latitudes ($b>20^\circ$) as a function of $S_{10}$ flux.} Blue line marks the $S90$ flux in our analysis, while black dashed line marks the corresponding value from \citep{DiMauro:2017ing}. {\it Right:} The gray histogram represents the numbers of AGNs and PSRs from the 4FGL catalog, as a function of the $S_1$ source flux. Red dots represent the source numbers from our F0 test catalog. Black dots represent sources that are recovered from F0 by our algorithm, accounting for true and false positives (TP+FP). Blue dots represent only TP. Notice that we got rid of FP by using the classification algorithm (see Sec. \ref{sec:classification}). {Vertical lines are drawn to guide the eye and represent proxies for detection thresholds  in our analysis (blue) and 4FGL (black, dashed)}. }\label{fig:temp2}
\end{figure*}

A rigorous comparison of the performance of our algorithms with those used to create the LAT catalogs is unfortunately out of the scope of the current work.  Such comparison requires the use of identical statistical criteria for the inclusion of sources and identical data sets. Fulfilling both of these criteria is not possible in this iteration of our work {because our simulated data is not compatible with that required for the algorithm used to create the catalog.  Nevertheless we qualitatively compare the algorithms below to offer some insights on the performance differences.

A very useful baseline for a qualitative comparison is provided by \cite{DiMauro:2017ing}, which uses Monte Carlo simulations of the gamma-ray sky, to calculate the detection efficiency (compatible with our definition of Completeness) of {a maximum likelihood} pipeline to detect point sources, making their results comparable with our approach. While our pipeline does not output the significance of the results in terms of a Test Statistic (TS), that work demonstrates that the detection efficiency varies only marginally when different TS cuts are applied (at least  in the explored range, from TS=16 to TS=25). We observe in Fig.~\ref{fig:temp2} (left), that our S90 flux threshold is $\sim 25\%$ better than the detection threshold in \cite{DiMauro:2017ing}, defined as the detection efficiency tends to one. There are, however, important caveats:  \cite{DiMauro:2017ing} uses 7 years of Pass 8 data (from 2008 August 4 to 2015 August 4), while we use 9.5 yrs of data. They also consider an energy range from 10 to 1000 GeV in  their analysis, while we train  our algorithm on the data set starting at energies $>500$ MeV (though we do not take the full advantage of the angular resolution of the LAT, unlike \cite{DiMauro:2017ing}).  The definition of the flux threshold is also slightly different.  Most of these choices can improve the performance of our algorithm, and are (at least in part) responsible for the observed lower detection threshold by our algorithm. 

In  terms  of the data set our choice is closer  to that  of the 4FGL catalog, which uses 8 years of data in the 50 MeV - 100 GeV range. One  way to quantify the source  detection threshold, introduced in the 3FGL catalog (\cite{Acero:2015gva}, Section 6.2.) is to plot the luminosity function  of  detected sources and to identify the  source detection threshold with the flux value at which the  luminosity function shows  departures from the power law  observed at high fluxes. We show the AGN+PSR number counts vs source flux in Fig.~\ref{fig:temp2}, right and see that a detection threshold defined in this way for our subset of 4FGL sources, corresponds to $S_1\sim 5\times 10^{-10}$ photons cm$^{-2}$ s$^{-1}$. Indeed, the 4FGL paper quotes as a source detection threshold derived in a similar way,  for all sources, as $2\times 10^{-12}$ erg cm$^{-2}$ s$^{-1}$, for energy flux above 100 MeV, which translates to a photon flux of $S_1\sim 2\times 10^{-10}$ photons cm$^{-2}$ s$^{-1}$ (assuming a spectral index of $-2$), consistent with our estimate above.  

We show the luminosity function for the sources in our test set (red dots, in Fig.~\ref{fig:temp2} right), together with sources detected by our algorithm (black dots in the figure). The detection threshold in this  case  can  be identified  with the flux value in which the two function depart (shown with a blue vertical line in Fig.~\ref{fig:temp2} right), the value that indeed overlaps with our definition of S90 when calculated for the whole sky.
{While there are again caveats in terms of exact dataset (8 vs. 9.5 years, >50 MeV vs. 500 MeV), the figure demonstrates that the flux thresholds for our network and 4FGL analysis are comparable.}  
{The global results of our approach also have the benefit of being nearly independent of the employed background model as demonstrated in the lower panels of Fig.~\ref{fig:temp}.  A more detailed comparison between the result using B1, B2, and B3 shows that around 90\% of the TP sources are common between each pair of results, while less than 30\% of the FP sources are common.  This can be contrasted with a comparison of the FL8Y\footnote{\url{https://fermi.gsfc.nasa.gov/ssc/data/access/lat/fl8y/}} and the 4FGL.  The former uses the B3 background model and has 5523 sources, while the latter uses the B1 background model and has 5064 sources, a global difference of around 10\%.  In addition, 215 sources in the 4FGL are flagged with analysis flag 1 and 2, which indicates that the source went below the 4FGL detection threshold when using B3 as the background model.  Our approach is therefore globally more stable against changes in the background model, while the fraction of overlap sources is similar.}

\begin{figure*}[h]
\begin{center}
\includegraphics[scale=0.35]{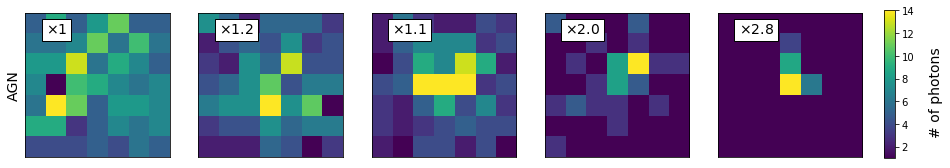}
\includegraphics[scale=0.35]{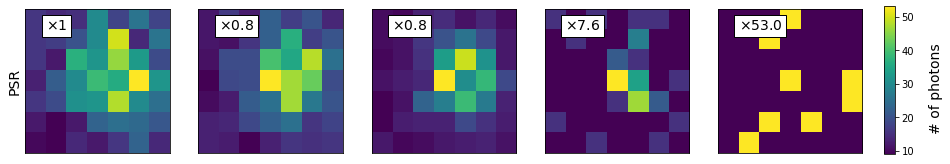}
\includegraphics[scale=0.35]{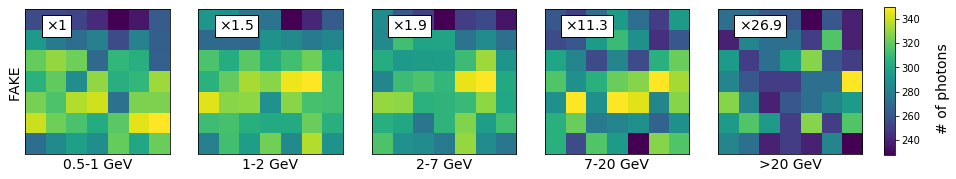}
\end{center}
\caption{From left to right we show the map of photon number count for each energy bin {for an example source of each class in each row. The color scale for each point source class is normalized to the maximum count of the first energy bin and we write in text box the level of amplification necessary for higher energy bins.} Top panel shows AGN source class which has a spectrum extending to the highest energy while the PSR (middle panel) cuts off at energies of few GeV. The bottom panel shows an example of a FAKE source.}
\label{fig:classification_sourcepatches}
\end{figure*}

\section{Point source classification}
\label{sec:classification}

Using deep neural networks it is not only possible to detect points sources, but these methods can also be used for classification of sources  by their type. In this work, we focused on point sources from AGN and PSR source classes, based only on {a small section of the simulated data extracted around their predicted central location, see Fig.~\ref{fig:classification_sourcepatches} and below.}

\subsection{Training data and network structure}

To train the classification network, we constructed a dataset based on the output of the point source detection algorithm. Namely, we selected a 7x7 pixel patch in 5 energy bins around the predicted point source.

We also required that the predicted position indeed contain a 7x7 box around in order to generate the point source patch. Thus, if there is a TP match where the predicted position is too close to the patch's edge, we dismissed this source for the classification training. We note that the absolute position on the sky was not part of the input labels. 

We used three labels for training: AGN and PSR, plus a new class, called 'FAKE' (which represents false positives, as identified by the point source detection network), see Fig.~\ref{fig:classification_sourcepatches}. The data set defined in  this way contains 50705 AGN images, 6269 PSR images and 4943 FAKE images. The large discrepancy in  the representation of  different  source classes presents a  challenge for the classification network, and therefore an additional step of ``balancing"  of the data set is needed. 

\begin{figure*}[h]
\begin{center}
\includegraphics[scale=0.4]{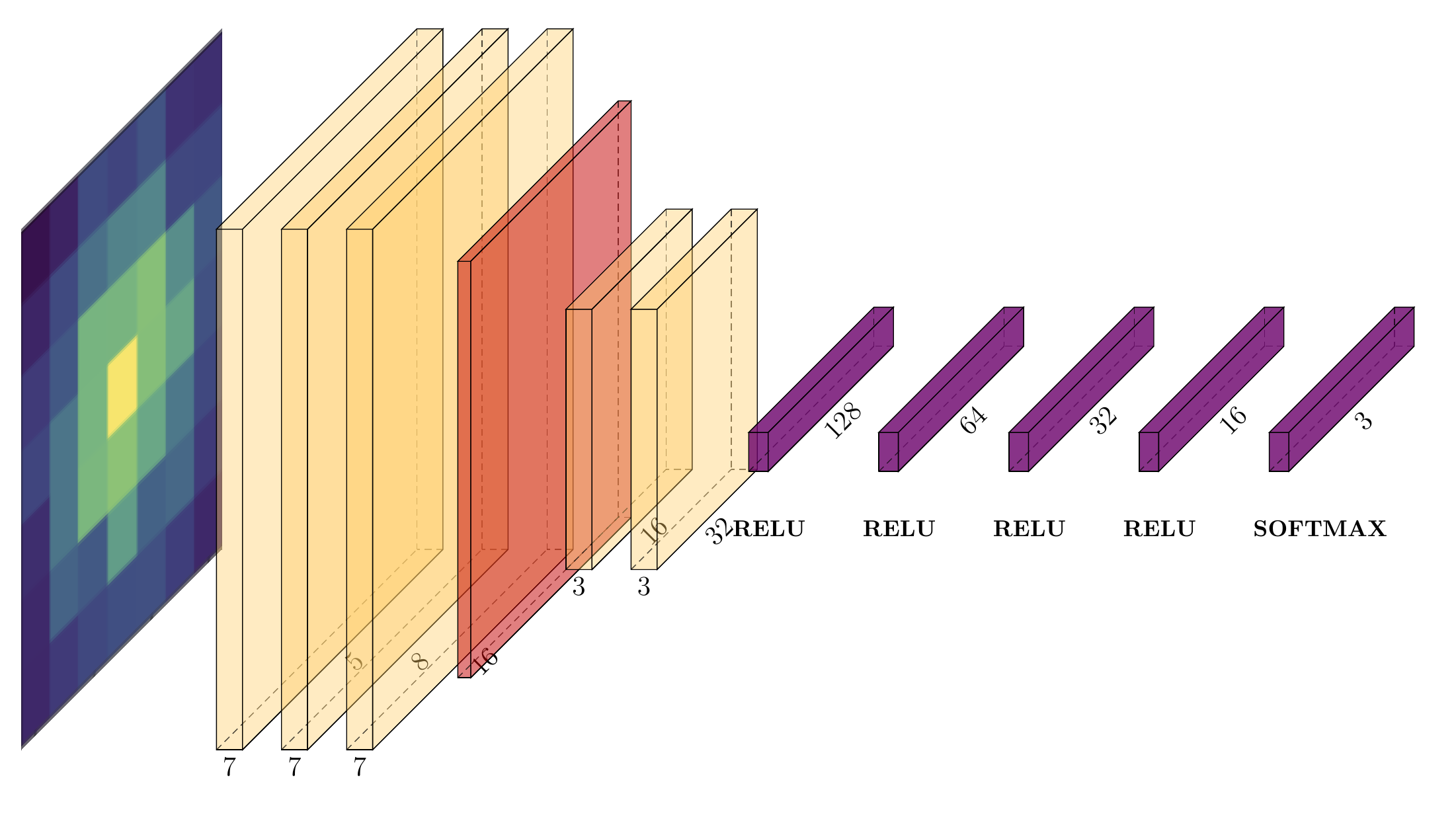}
\end{center}
\caption{Visualization of the classification network. The first layer is a batch normalization layer \citep{paper_batch_normalization} on a 7x7 image with 5 energy bins. Next are four convolutional layers and a max-pool layer and then five dense layers that make an output prediction.}
\label{fig:classification_nn}
\end{figure*}

To fix the unbalanced representation, 20000 AGN images are selected randomly, while random images from the PSR and FAKE classes are duplicated to create equal distributions. Since we apply Poisson noise to the Asimov data set {\it a posteriori}\footnote{The  original images represent the Asimov dataset, so during training Poisson samples are taken.}, every sample is still a unique image even though we have multiple copies of the same Asimov image in the dataset during training. In addition, the data set is `normalized' in order to  facilitate the network application. Namely, the Poisson sampled images are preprocessed pixel by pixel in every energy bin by substracting the mean and dividing by the standard deviation of the specific pixel across the whole dataset. 

\begin{figure*}[h]
\begin{center}
\includegraphics[scale=0.45]{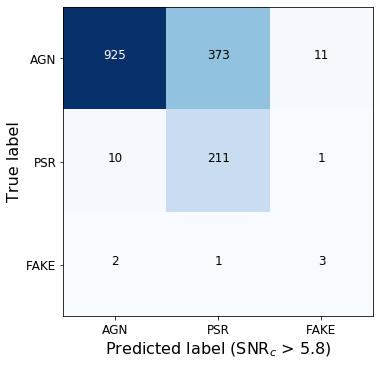}
\includegraphics[scale=0.44]{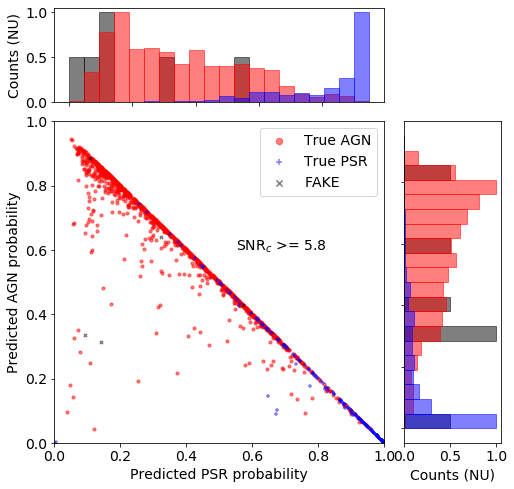}\\
\includegraphics[scale=0.45]{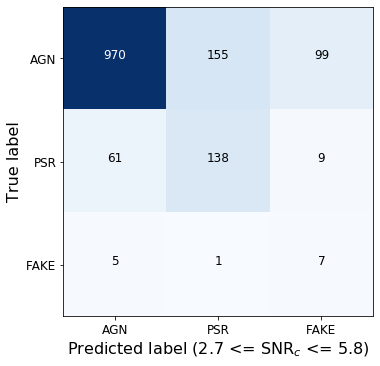}
\includegraphics[scale=0.44]{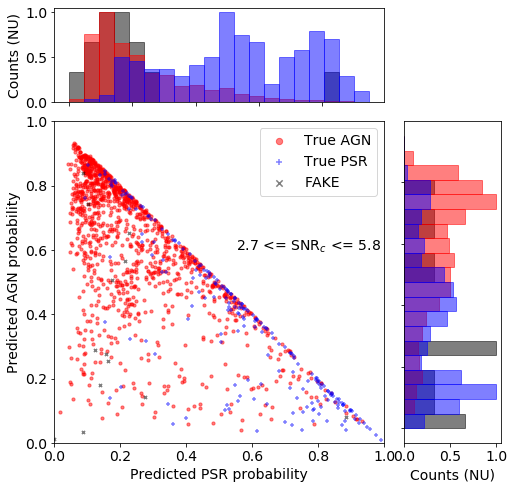}\\
\includegraphics[scale=0.45]{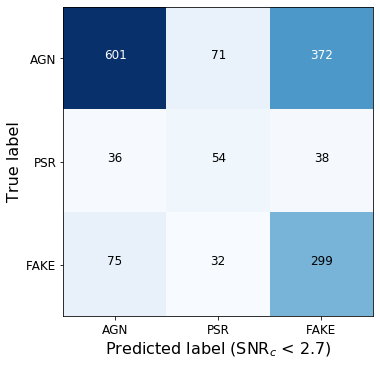}
\includegraphics[scale=0.44]{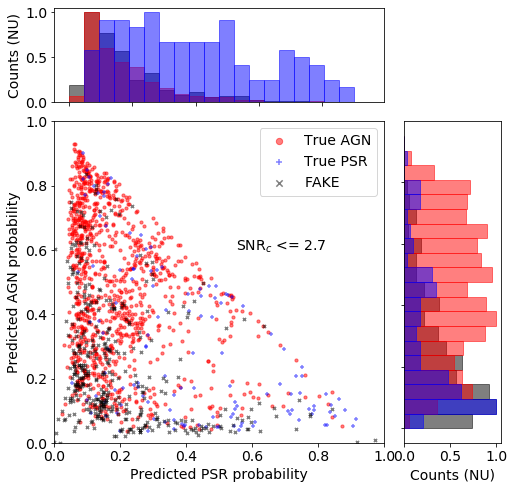}
\end{center}
\caption{{Confusion matrix as a function of  SNR$_c$ (left). Predicted probability scatter plot for the same SNR$_c$ sample (right). See text for details.}}
\label{fig:confusionmatrix-snr}
\end{figure*}

We trained a deep neural classification network with the architecture visualized in Fig.~\ref{fig:classification_nn} to classify the sources. All activation functions are Rectified Linear Units (ReLUs) \citep{relu}, except for the last layer which is a softmax layer. The optimizer is the Adam \citep{adam_optimizer} optimizer and the loss function is a categorical cross entropy. The network is trained with an unbounded number of epochs, but the learning rate is halved if the loss has not improved for 5 consecutive epochs and learning is stopped when the loss has not improved for 50 consecutive epochs. 

In order to make a safety check that the approach used to balance the class distributions (balanced approach) indeed works better than the approach considering the original distribution of classes (unbalanced approach), we compared the isolated performance of two classification networks, one trained with the balanced sample and the other one trained with the unbalanced one. For the evaluation, we considered a fixed test data sample (F0-B1) that contains the classes following the original distribution (as it would appear after the localization algorithm.). The Completeness for the balanced approach is similar for each class (70\%, 70\%, 75\%) for (AGN,  PSR, FAKE), while the unbalanced case systematically prefers AGNs with the following distribution (99\%, 30\%, 13\%). We focused on the evaluation of Completeness because we want to find the classification network which is able to identify the correct source type with a probability that is independent of the source class. In this sense we clearly see that the balanced network does a better job than the unbalanced one.  

\subsection{Results of PS classification algorithm and related uncertainties}

We illustrate the performance of the classification network (trained on data from the UNEK algorithm) in Fig.~\ref{fig:confusionmatrix-snr}. Three panels, from top to bottom, show the results for sources with the  highest  signal-to-noise (SNR>5.8)\footnote{The FAKE source patches are defined as $7\times7\times5$ cubes around FP predicted positions and we define the SNR parameter as the ratio between the number of photons from the source layer with the higher number of counts (either PSR or AGN) with the highest value and the square root of the sum of IEM and source layers, where the number of photons is evaluated at the predicted position of the FP source. Since the predicted FAKE sources do not have a true source match the value of SNR is expected to be small.} and the faintest sources (SNR<2.7) in the bottom row\footnote{SNR binning is chosen so that there is approximately the same number of  sources in each bin.}. Left panels show total numbers  of  sources in  the ``True label" vs. ``Predicted Label" (i.e. ``confusion") matrix. Note that the entries in the confusion matrices are the highest predicted class probabilities. We observe that the matrix is largely diagonal, as expected for a high performance algorithm. In the right panels, we visualize the performance in Predicted AGN label vs Predicted  PSR label plane, demonstrating that sources  nicely cluster in three regions, AGNs at the upper left, PSRs at the lower right and FAKEs at the low source probability corners, {in all three SNR bins}. For low SNR (lowest panels) this distinction is {somewhat} less pronounced, while the number  of FAKE sources  is  increased, as expected.

\begin{figure*}[h]
\begin{center}
\includegraphics[scale=0.48]{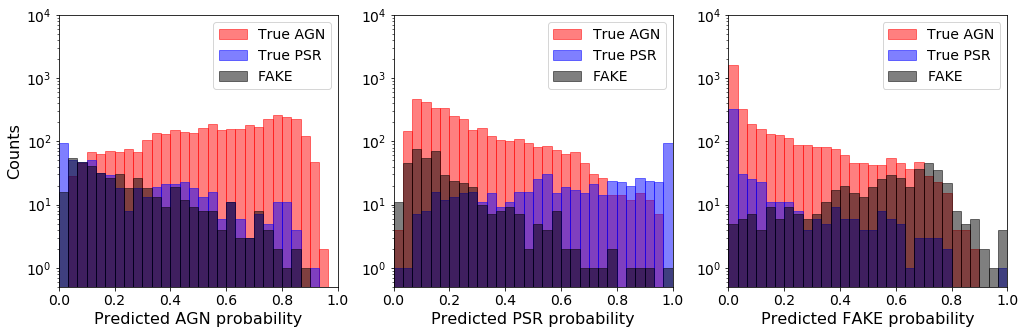}
\end{center}
\caption{{Distribution of the number of point sources as a function of the class probability. Evaluation of CNN classification model. The model is trained with the predictions of UNEK over 10k patches (F0-B1). We use Poisson repetitions in order to balance the distribution of source classes. We use one single test sample, obtained from 768 patches of catalog 235 (F0-B1). See text for details.}}
\label{fig:threesourceclasses}
\end{figure*}

In Fig.~\ref{fig:threesourceclasses} we show the distribution of the detected sources, for the three source classes, as a function of predicted AGN probability (left column), predicted PSR probability (middle) and predicted  FAKE probability (right column), for all SNR values. This demonstrates that by imposing a given cut on the prediction probability one can define predicted samples of a desired Purity. One can also note that PSRs have lower classification Purity than AGNs, which is likely in part due to the fact that they are found mostly  in the Galactic plane, where astrophysical backgrounds are higher. 

\begin{figure*}[ht]
\begin{center}
\includegraphics[scale=0.62]{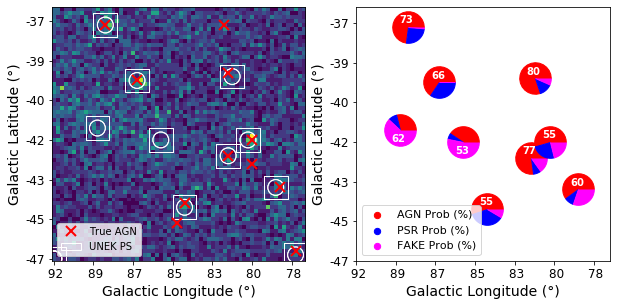}\\
\includegraphics[scale=0.62]{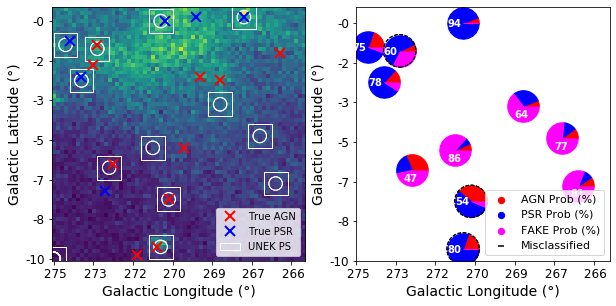}
\end{center}
\caption{{\it Left:} results of the point source detection by UNEK algorithm (white circles), overlaid with the original gamma-ray map in the  0.5-1 GeV energy bin. The cross marks true sources in the test data, with the color representing their true class. {White squares mark the 7x7 pixel region that is used for source classification}. {\it Right} Results of the classification  algorithm are presented with colored circles, where the color represents the probability of a given predicted class. We also explicitly mark the highest probability of a given class, as output by  the network. Black-dashed circle edges mark miss-classified sources. Note that the sources that appear at  the edges of a patch on the left are not included in the classification procedure and are therefore missing in the right figure. The {\it Top} row focuses on high latitudes, while the  {\it Bottom} row shows a region close to the Galactic plane.}
\label{fig:classification-example}
\end{figure*}

Finally, two example patches of the sky (chosen at  high- (top panel) and low- (lower panel) latitudes), together with the results of our localization and classification algorithm, are shown in Fig.~\ref{fig:classification-example}. In particular, left panels show the sky image, {with true sources marked with crosses},  overlaid with point sources detected by our algorithm, ({where} {white circles show our localization radius, while white squares show the region of the sky around the source that we use for classification}). Those sources are  then fed to the classification  algorithm and the resulting probability of the source class is  shown in the right panels. We mark misclassified sources with a black dashed border  and notice that the number of missclassified as well as nondetected sources is higher in the patch that is closer to the Galactic plane (bottom row), as expected.

In terms of uncertainties, in  this work we adopted a deterministic approach for the predictions of the point source locations and softmax probabilities for the classification of these point sources (a vector of class prediction probabilities per point source). While this does provide some degree of certainty to a particular prediction, the inherent uncertainty from the model are not embedded explicitly into this prediction. This is, however, not an  inherent limitation of our network architecture and model uncertainties (i.e. epistemic error) could be included in our pipeline in the following way. 
As the network architecture and training dataset size are never perfect, training multiple networks will yield slightly different results. By applying a dropout layer after every hidden layer with weights (which are not removed after training), every inference will have a slightly different network architecture as certain weights are turned off. One can run the same input $N$ times through the network, every time with a different randomized set of weights that are turned off, to yield $N$ predictions. The mean and standard deviation of the predictions can then be simply calculated from these predictions to obtain a network uncertainty of the prediction. A mathematical derivation of the epistemic error can be found at \citep{NIPS2017_7141_bayesian_deep_learning}. When a network is very accurately trained, these different inferences will yield similar results. However, when a network is trained with very limited data or when the network architecture is not good enough to make accurate predictions, slight variations in weight configuration will yield vastly different results. {As stated above, we neglect the epistemic uncertainty in this iteration  of our project, but plan to incorporate it in the follow up, when we apply the network to  the real data.}

\section{Summary}

In this work we set up an automatic deep learning image segmentation pipeline which localizes and classifies gamma-ray point sources starting from the raw Fermi-LAT data. The data used to train these models includes hundreds of simulated sky instances based on the 4FGL Fermi-LAT catalog, that contain time integrated photon number counts in pixels that represent spatial and energy space. The simulations considered the contributions of point sources and diffuse emission.

We performed the  task in two steps: In a first step, we run the PS finding algorithms (UNEK  and CNET) and demonstrate that sources are found with the Purity and Completeness higher than $90\%$  (the so-called S90 score) {down to SNR=3 (and SBR=1)}. In terms of the flux sensitivity, the performance of the PS finding algorithms is at  least  as good as that of the traditional 4FGL pipeline, while  having a strong advantage of being robust with  respect to the diffuse emission across the range of models we tested.

In a second step we focused on source classification and trained a deep neural network to output class probabilities. While we adopted only three source classes and use only spectral information for classification, the algorithm is clearly capable of separating the sources. As expected, PSRs prove harder to classify, due to their localization in high background regions. 

In summary, our work demonstrates that image segmentation and classification neural networks provide a powerful tool to detect and classify point  sources from gamma-ray images of the sky. We release our training data and network scripts on github {and invite the community to participate in a challenge, with the goal of improving the current performance.}

\clearpage

\begin{acknowledgements}
R. RdA acknowledges partial funding/support from the Elusives ITN (Marie Sk\l{}odowska-Curie grant agreement No 674896), the ``SOM Sabor y origen de la Materia" (FPA 2017-85985-P). G.~Z.~and C.~E. acknowledge the financial support from the Slovenian Research Agency (grants P1-0031, I0-0033, J1-1700 and the Young Researcher  program). B.~P. acknowledges the financial support from the joint committee ESO-Government of Chile (Postdoctoral Fellowship). The idea for this project started at the annual DarkMachines meeting in Trieste and project was subsequently developed, in part, within the DarkMachines Point Source working group. We are grateful for  constructive comments we received from Jean Ballet, Tobby Burnet, Seth Digel and Benoit Lott.
\end{acknowledgements}

\bibliographystyle{aa}
\bibliography{PSwCN}

\appendix

\section{Parameters for the synthetic catalogs}
\label{app:parameter_plots}

In this section we provide a comparison of the {distributions of spectral parameters in the 4FGL to those of a} synthetic catalog generated by the reasoning described in Sec.~\ref{sec:training_data_generation}.  In particular, we compare the distributions of the main catalog parameters for {the AGN population} in Fig.~\ref{fig:agn_properties} and for the PSR population in Fig.~\ref{fig:psr_properties}.  By design the distributions of our synthetic catalogs match those of the 4FGL very well.

In addition, we give details about the luminosity function sampling, which follows two general strategies:

{\it `Flat' extrapolation:}
In this case we assume that the luminosity function is hard, resulting in most of the sources being above (or just below) the source detection threshold. This  strategy results in a relatively quick mock data generation, as the number of sub-threshold sources is lower than in the power-law extrapolation described below. We therefore use this set-up for training. In particular, we follow the algorithm below:
\begin{enumerate}
    \item We bin the 4FGL AGN (PSR) sources according to their listed $E_{100}$ values and prepare a list of mock sources with the very same bin width (including additional bins down to the lower threshold reported above).
    \item For each instance of a mock AGN (PSR) catalog, we create two sets of random numbers:
    
    \begin{itemize}
        \item[(a)] $N_{\mathrm{min}}\in\left[50,250\right]$ ($N_{\mathrm{min}}\in\left[15,25\right]$), representing the number of sources inside the lowest energy flux bin of the mock catalog {for AGN (PSR)} and
        \item[(b)] a list $N_{\mathrm{noise}}$ of random numbers drawn from a uniform distribution with values between 0.8 and 1.3 whose length corresponds to the number of energy flux bins of the original 4FGL distribution of AGN (PSR) $\phi^{\rm E}$ to the right of the peak of $\mathrm{d}N/\mathrm{d}S$ (c.f.~Figs.~\ref{fig:agn_properties} or \ref{fig:psr_properties}, respectively, and the blue histogram for a visualization of the distribution in question).  
        \item [(c)] We extend the synthetic catalog to energy fluxes $F^{E, TH}_{FL}$ as small as $3.4\times10^{-13}\;\mathrm{erg}$ $\mathrm{cm}^{-2}$ $\mathrm{s}^{-1}$, which is more than one order of magnitude below the 4FGL flux threshold of $F^{E, TH}_{{4FGL}}\simeq 2\times10^{-12}\;\mathrm{erg}\,\mathrm{cm}^{-2}\,\mathrm{s}^{-1}$ at high latitudes $|b|>10^{\circ}$ \citep{Fermi-LAT:2019yla}. 
    \end{itemize}
    
    \item We draw the spectral and spatial parameters of a candidate mock AGN  (PSR) source, compute the resulting value of $E_{100}$ and determine the energy flux histogram bin $k$ it would contribute to. A source is accepted if the number of entries in the respective bin does not overshoot the number of 4FGL sources times $\left(N_{\mathrm{noise}}\right)_k$ in the original 4FGL histogram or, in case the candidate source in question would fall into the lowest energy flux bin, the number of sources there is lower than $N_{\mathrm{min}}$. If a bin in the original 4FGL distribution contains zero or one source, our synthetic catalog may maximally populate these bins with two sources.
    
    \item The synthetic catalog generation is completed when the lowest $E_{100}$ bin contains $N_{\mathrm{min}}$ entries and all bins above the peak of 4FGL's $\mathrm{d}N/\mathrm{d}S$ have been fully populated according to the upper threshold determined by $N_{\mathrm{noise}}$.
\end{enumerate}

\noindent{\it `Power-law' extrapolation:}
In order to check the robustness of our pipeline to the luminosity function variations, we prepare a {\em single} whole sky map based on a power law extrapolation of the original 4FGL luminosity function. To this end, we performed a linear regression of the binned $F^E_{100}$ distribution of all 4FGL AGN sources with respect to energy fluxes above the peak value. In $\log{N}$-$\log{E_{100}}$ space, the best-fit result of the linear regression finds for the slope a value of $m = -0.96$ while the interception point with the x-axis is located at $x_0 = -8.64$. Consequently, we have populated the bins of the energy flux mock histogram with the expected number of entries according to the linear regression down to an energy flux of $F^{E, TH}_{PL}=0.01~\Phi^{\rm E}_{\rm 4FGL}$= $2.2\times10^{-14}\;\mathrm{erg}\,\mathrm{cm}^{-2}\,\mathrm{s}^{-1}$. The selection of spectral and spatial parameters still follow the prescription of the alternative algorithm. We stress that having a mock  catalog that has numerous (overlapping) faint sources, 
could potentially lower the localization sensitivity and therefore provides an important test for our algorithm. 

\begin{figure*}[h!]
\begin{center}
\includegraphics[width=0.85\linewidth]{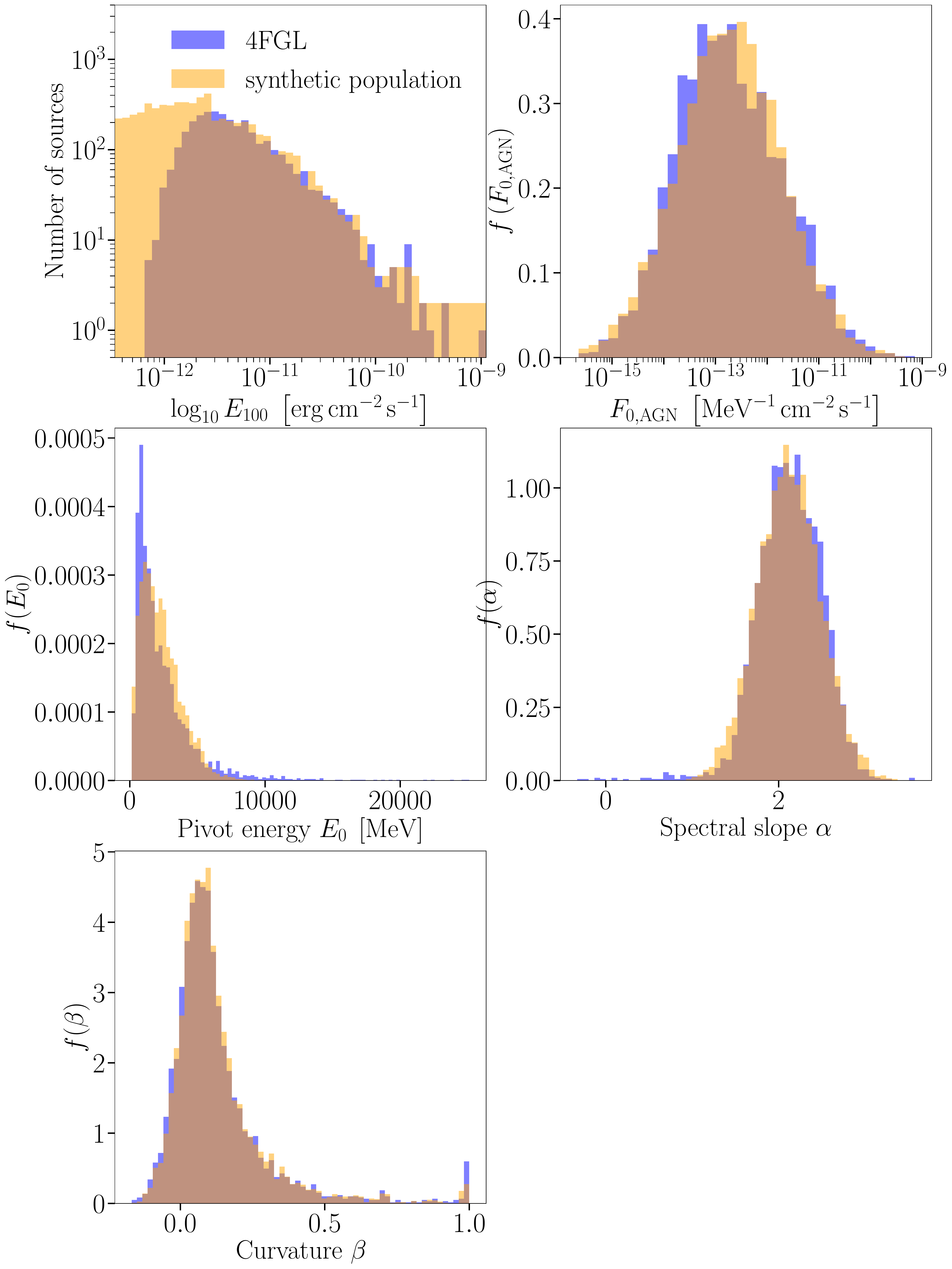}
\end{center}
\caption{Comparison of the mock catalog's (orange) and 4FGL's (blue) AGN properties with respect to four spectral parameters and the energy flux $E_{100}$ within the range from 0.1 to 100 GeV. The latter quantity is displayed in the upper left panel, while the flux density  $F_{0,\mathrm{AGN}}$ is shown next to it. The remaining log-normal parameters are the pivot energy $E_0$ (middle left panel), spectral slope $\alpha$ (middle right panel) and the curvature $\beta$ (lower left panel). The y-axis of the $E_{100}$-distributions is given in actual object numbers whereas the other four panels show the normalized distributions of the respective parameters. \label{fig:agn_properties}}
\end{figure*}

\begin{figure*}[h!]
\begin{center}
\includegraphics[width=0.85\linewidth]{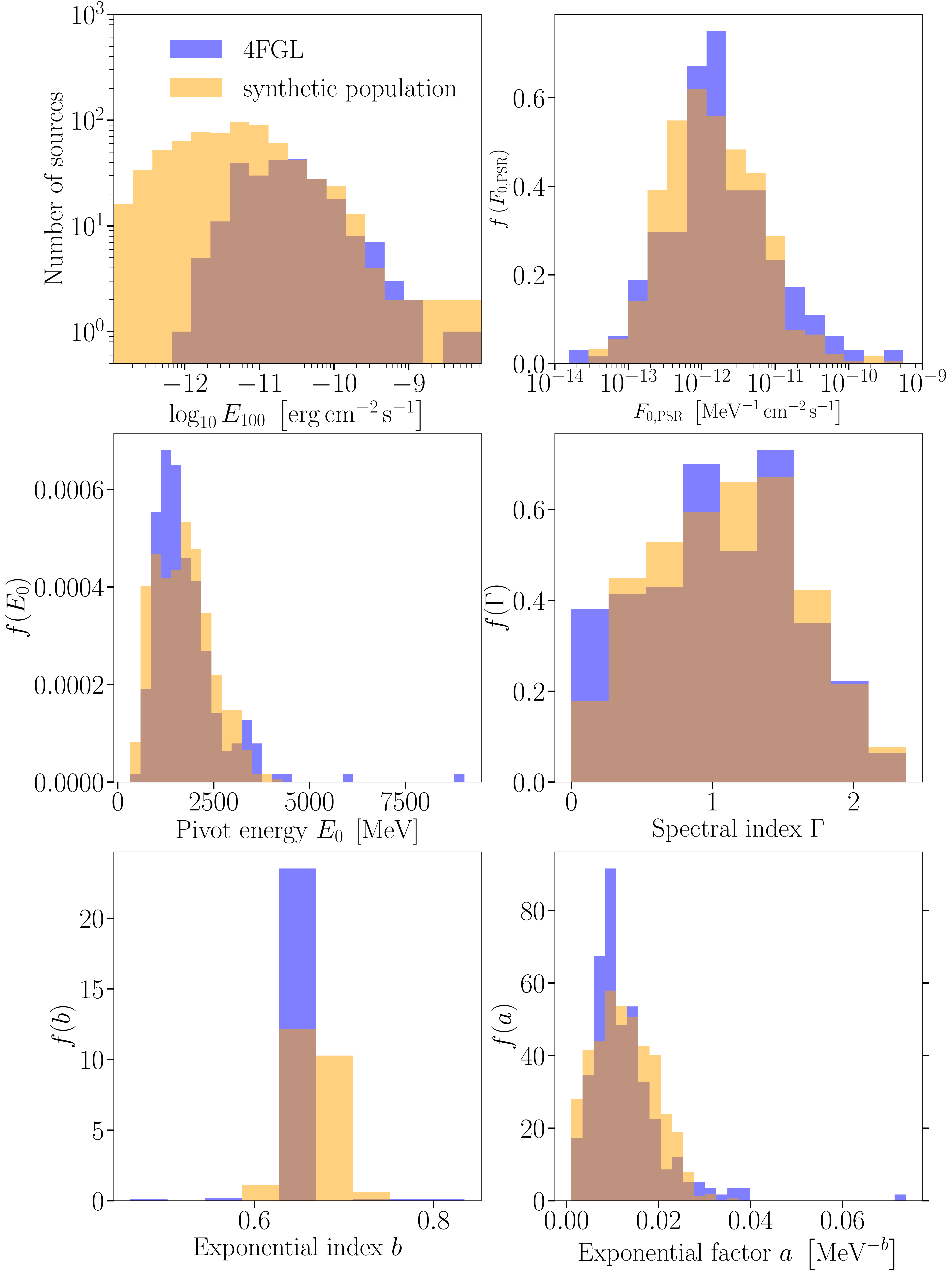}
\end{center}
\caption{Comparison of the mock catalog's (orange) and 4FGL's (blue) pulsar properties with respect to four spectral parameters and the energy flux $E_{100}$ within the range from 0.1 to 100 GeV. The latter quantity is displayed in the upper left panel, while the flux density  $F_{0,\mathrm{PSR}}$ is shown next to it. The remaining PLEC parameters are the pivot energy $E_0$ (center left panel), the low-energy spectral slope $\Gamma$ (center right panel), the exponential index $b$ (lower left panel) and the exponential factor $a$ (lower right panel).
The y-axis of the $E_{100}$-distributions is given in actual object numbers whereas the other four panels show the normalized distributions of the respective parameters.   \label{fig:psr_properties}}
\end{figure*}

\section{UNEK clustering algorithm}
\label{app:UNEKalgorithm}

In Alg.~\ref{alg:kmeans-based} we show the pseudo code of the $k$-means based algorithm, which is applied in the second step of the UNEK model. Hyper-parameters are fixed to baseline values that where used to compute the performance of the UNEK along this paper. 

\begin{algorithm}[ht]
 \KwData{first layer of U-Net output}
 \KwResult{list of $k$-means centroids corresponding to $k_\text{best}$}
 $D\leftarrow$ copy Data\;
 $l_{sth} \leftarrow$ define the label score threshold\;
 $l_{sng}\leftarrow$ define the penalization factor\;
 $R\leftarrow$ define the radius of the region of interest\;
 $V_D \leftarrow$ from $D$, get the pixel positions with label score greater than $l_{sth}$\;
 $k_{\text{best}} \leftarrow 0$\;
 $S_{k,\text{max}} \leftarrow 0$\;
 \For{k in range(1,50)}{
 $k$-$centroids \leftarrow$ run $k$-means on $V_{D}$, for $k$ clusters\;
 $S_k \leftarrow 0$\;
  \For{c in k-centroids}{
  $P_c \leftarrow$ find the pixels of $D$ inside a radius $R$ around $c$\;
  $S_k\leftarrow$ considering the pixels of $P_c$, update the sum of scores\;
 $D\leftarrow$ redefine the scores of the pixels in $P_c$ with a value $l_{sng}$\;
  }
  \If{$S_k > S_{k,\text{max}}$}{
  $k_\text{best} \leftarrow k$\;
  $S_{k, \text{max}} \leftarrow S_k$\;
  }
 }
\
\caption{$k$-means based clustering algorithm. The default algorithm uses $l_{sth}=0.2$, $l_{snn}=-10$ and $R=5$ pixels.}
\label{alg:kmeans-based}
\end{algorithm}

\end{document}